\documentclass[longauth]{aaEC}

\usepackage{graphicx}
\usepackage{natbib}
\usepackage{scalerel}
\usepackage{lastpage}

\usepackage[nolist,nohyperlinks]{acronym}

\usepackage[table]{xcolor}
\usepackage{mathtools} 
\usepackage{dcolumn}

\usepackage{txfonts}
\usepackage[pdfencoding=auto,psdextra]{hyperref}
\hypersetup{
    colorlinks=true,
    linkcolor=blue,
    filecolor=magenta,      
    urlcolor=blue,
    citecolor=blue
}
\makeatletter
\renewcommand*\aa@pageof{, page \thepage{} of \pageref*{LastPage}}
\makeatother

\usepackage[utf8]{inputenc}

\usepackage[switch, modulo]{lineno}

\usepackage{euclid}
\usepackage[normalem]{ulem}

\usepackage[nameinlink,capitalise]{cleveref}
\crefname{section}{Sect.}{Sects.}
\Crefname{section}{Section}{Sections}
\crefname{figure}{Fig.}{Figs.}
\Crefname{figure}{Figure}{Figures}
\crefname{equation}{Eq.}{Eqs.}
\Crefname{equation}{Equation}{Equations}
\crefname{table}{Table}{Tables}
\crefname{appendix}{Appendix}{Appendices}

\begin{document}

\title{Euclid}
\subtitle{VI. NISP-P optical ghosts}    

\acrodef{EWS}{{\em Euclid} Wide Survey}
\acrodef{PV}{Performance Verification}
\acrodef{HST}{Hubble Space Telescope}

\newcommand{\orcid}[1]{} 
\author{Euclid Collaboration: K.~Paterson\orcid{0000-0001-8340-3486}\thanks{\email{paterson@mpia.de}}\inst{\ref{aff1}}
\and M.~Schirmer\orcid{0000-0003-2568-9994}\inst{\ref{aff1}}
\and K.~Okumura\inst{\ref{aff2}}
\and B.~Venemans\orcid{0000-0001-9024-8322}\inst{\ref{aff3}}
\and K.~Jahnke\orcid{0000-0003-3804-2137}\inst{\ref{aff1}}
\and N.~Aghanim\orcid{0000-0002-6688-8992}\inst{\ref{aff4}}
\and B.~Altieri\orcid{0000-0003-3936-0284}\inst{\ref{aff5}}
\and A.~Amara\inst{\ref{aff6}}
\and S.~Andreon\orcid{0000-0002-2041-8784}\inst{\ref{aff7}}
\and C.~Baccigalupi\orcid{0000-0002-8211-1630}\inst{\ref{aff8},\ref{aff9},\ref{aff10},\ref{aff11}}
\and M.~Baldi\orcid{0000-0003-4145-1943}\inst{\ref{aff12},\ref{aff13},\ref{aff14}}
\and A.~Balestra\orcid{0000-0002-6967-261X}\inst{\ref{aff15}}
\and S.~Bardelli\orcid{0000-0002-8900-0298}\inst{\ref{aff13}}
\and P.~Battaglia\orcid{0000-0002-7337-5909}\inst{\ref{aff13}}
\and A.~Biviano\orcid{0000-0002-0857-0732}\inst{\ref{aff9},\ref{aff8}}
\and A.~Bonchi\orcid{0000-0002-2667-5482}\inst{\ref{aff16}}
\and E.~Branchini\orcid{0000-0002-0808-6908}\inst{\ref{aff17},\ref{aff18},\ref{aff7}}
\and M.~Brescia\orcid{0000-0001-9506-5680}\inst{\ref{aff19},\ref{aff20}}
\and J.~Brinchmann\orcid{0000-0003-4359-8797}\inst{\ref{aff21},\ref{aff22}}
\and S.~Camera\orcid{0000-0003-3399-3574}\inst{\ref{aff23},\ref{aff24},\ref{aff25}}
\and G.~Ca\~nas-Herrera\orcid{0000-0003-2796-2149}\inst{\ref{aff26},\ref{aff27},\ref{aff3}}
\and V.~Capobianco\orcid{0000-0002-3309-7692}\inst{\ref{aff25}}
\and J.~Carretero\orcid{0000-0002-3130-0204}\inst{\ref{aff28},\ref{aff29}}
\and S.~Casas\orcid{0000-0002-4751-5138}\inst{\ref{aff30}}
\and M.~Castellano\orcid{0000-0001-9875-8263}\inst{\ref{aff31}}
\and G.~Castignani\orcid{0000-0001-6831-0687}\inst{\ref{aff13}}
\and S.~Cavuoti\orcid{0000-0002-3787-4196}\inst{\ref{aff20},\ref{aff32}}
\and K.~C.~Chambers\orcid{0000-0001-6965-7789}\inst{\ref{aff33}}
\and A.~Cimatti\inst{\ref{aff34}}
\and C.~Colodro-Conde\inst{\ref{aff35}}
\and G.~Congedo\orcid{0000-0003-2508-0046}\inst{\ref{aff36}}
\and C.~J.~Conselice\orcid{0000-0003-1949-7638}\inst{\ref{aff37}}
\and L.~Conversi\orcid{0000-0002-6710-8476}\inst{\ref{aff38},\ref{aff5}}
\and Y.~Copin\orcid{0000-0002-5317-7518}\inst{\ref{aff39}}
\and F.~Courbin\orcid{0000-0003-0758-6510}\inst{\ref{aff40},\ref{aff41},\ref{aff42}}
\and H.~M.~Courtois\orcid{0000-0003-0509-1776}\inst{\ref{aff43}}
\and A.~Da~Silva\orcid{0000-0002-6385-1609}\inst{\ref{aff44},\ref{aff45}}
\and R.~da~Silva\orcid{0000-0003-4788-677X}\inst{\ref{aff31},\ref{aff16}}
\and H.~Degaudenzi\orcid{0000-0002-5887-6799}\inst{\ref{aff46}}
\and G.~De~Lucia\orcid{0000-0002-6220-9104}\inst{\ref{aff9}}
\and A.~M.~Di~Giorgio\orcid{0000-0002-4767-2360}\inst{\ref{aff47}}
\and J.~Dinis\orcid{0000-0001-5075-1601}\inst{\ref{aff44},\ref{aff45}}
\and H.~Dole\orcid{0000-0002-9767-3839}\inst{\ref{aff4}}
\and F.~Dubath\orcid{0000-0002-6533-2810}\inst{\ref{aff46}}
\and X.~Dupac\inst{\ref{aff5}}
\and S.~Dusini\orcid{0000-0002-1128-0664}\inst{\ref{aff48}}
\and A.~Ealet\orcid{0000-0003-3070-014X}\inst{\ref{aff39}}
\and S.~Escoffier\orcid{0000-0002-2847-7498}\inst{\ref{aff49}}
\and M.~Farina\orcid{0000-0002-3089-7846}\inst{\ref{aff47}}
\and R.~Farinelli\inst{\ref{aff13}}
\and F.~Faustini\orcid{0000-0001-6274-5145}\inst{\ref{aff16},\ref{aff31}}
\and S.~Ferriol\inst{\ref{aff39}}
\and F.~Finelli\orcid{0000-0002-6694-3269}\inst{\ref{aff13},\ref{aff50}}
\and S.~Fotopoulou\orcid{0000-0002-9686-254X}\inst{\ref{aff51}}
\and N.~Fourmanoit\orcid{0009-0005-6816-6925}\inst{\ref{aff49}}
\and M.~Frailis\orcid{0000-0002-7400-2135}\inst{\ref{aff9}}
\and E.~Franceschi\orcid{0000-0002-0585-6591}\inst{\ref{aff13}}
\and P.~Franzetti\inst{\ref{aff52}}
\and S.~Galeotta\orcid{0000-0002-3748-5115}\inst{\ref{aff9}}
\and K.~George\orcid{0000-0002-1734-8455}\inst{\ref{aff53}}
\and W.~Gillard\orcid{0000-0003-4744-9748}\inst{\ref{aff49}}
\and B.~Gillis\orcid{0000-0002-4478-1270}\inst{\ref{aff36}}
\and C.~Giocoli\orcid{0000-0002-9590-7961}\inst{\ref{aff13},\ref{aff14}}
\and J.~Gracia-Carpio\inst{\ref{aff54}}
\and B.~R.~Granett\orcid{0000-0003-2694-9284}\inst{\ref{aff7}}
\and A.~Grazian\orcid{0000-0002-5688-0663}\inst{\ref{aff15}}
\and F.~Grupp\inst{\ref{aff54},\ref{aff53}}
\and L.~Guzzo\orcid{0000-0001-8264-5192}\inst{\ref{aff55},\ref{aff7},\ref{aff56}}
\and S.~V.~H.~Haugan\orcid{0000-0001-9648-7260}\inst{\ref{aff57}}
\and H.~Hoekstra\orcid{0000-0002-0641-3231}\inst{\ref{aff3}}
\and W.~Holmes\inst{\ref{aff58}}
\and F.~Hormuth\inst{\ref{aff59}}
\and A.~Hornstrup\orcid{0000-0002-3363-0936}\inst{\ref{aff60},\ref{aff61}}
\and P.~Hudelot\inst{\ref{aff62}}
\and M.~Jhabvala\inst{\ref{aff63}}
\and E.~Keih\"anen\orcid{0000-0003-1804-7715}\inst{\ref{aff64}}
\and S.~Kermiche\orcid{0000-0002-0302-5735}\inst{\ref{aff49}}
\and A.~Kiessling\orcid{0000-0002-2590-1273}\inst{\ref{aff58}}
\and R.~Kohley\inst{\ref{aff5}}
\and B.~Kubik\orcid{0009-0006-5823-4880}\inst{\ref{aff39}}
\and M.~K\"ummel\orcid{0000-0003-2791-2117}\inst{\ref{aff53}}
\and M.~Kunz\orcid{0000-0002-3052-7394}\inst{\ref{aff65}}
\and H.~Kurki-Suonio\orcid{0000-0002-4618-3063}\inst{\ref{aff66},\ref{aff67}}
\and A.~M.~C.~Le~Brun\orcid{0000-0002-0936-4594}\inst{\ref{aff68}}
\and D.~Le~Mignant\orcid{0000-0002-5339-5515}\inst{\ref{aff69}}
\and S.~Ligori\orcid{0000-0003-4172-4606}\inst{\ref{aff25}}
\and P.~B.~Lilje\orcid{0000-0003-4324-7794}\inst{\ref{aff57}}
\and V.~Lindholm\orcid{0000-0003-2317-5471}\inst{\ref{aff66},\ref{aff67}}
\and I.~Lloro\orcid{0000-0001-5966-1434}\inst{\ref{aff70}}
\and G.~Mainetti\orcid{0000-0003-2384-2377}\inst{\ref{aff71}}
\and D.~Maino\inst{\ref{aff55},\ref{aff52},\ref{aff56}}
\and E.~Maiorano\orcid{0000-0003-2593-4355}\inst{\ref{aff13}}
\and O.~Mansutti\orcid{0000-0001-5758-4658}\inst{\ref{aff9}}
\and S.~Marcin\inst{\ref{aff72}}
\and O.~Marggraf\orcid{0000-0001-7242-3852}\inst{\ref{aff73}}
\and K.~Markovic\orcid{0000-0001-6764-073X}\inst{\ref{aff58}}
\and M.~Martinelli\orcid{0000-0002-6943-7732}\inst{\ref{aff31},\ref{aff74}}
\and N.~Martinet\orcid{0000-0003-2786-7790}\inst{\ref{aff69}}
\and F.~Marulli\orcid{0000-0002-8850-0303}\inst{\ref{aff75},\ref{aff13},\ref{aff14}}
\and R.~Massey\orcid{0000-0002-6085-3780}\inst{\ref{aff76}}
\and H.~J.~McCracken\orcid{0000-0002-9489-7765}\inst{\ref{aff62}}
\and E.~Medinaceli\orcid{0000-0002-4040-7783}\inst{\ref{aff13}}
\and S.~Mei\orcid{0000-0002-2849-559X}\inst{\ref{aff77},\ref{aff78}}
\and M.~Meneghetti\orcid{0000-0003-1225-7084}\inst{\ref{aff13},\ref{aff14}}
\and E.~Merlin\orcid{0000-0001-6870-8900}\inst{\ref{aff31}}
\and G.~Meylan\inst{\ref{aff40}}
\and A.~Mora\orcid{0000-0002-1922-8529}\inst{\ref{aff79}}
\and M.~Moresco\orcid{0000-0002-7616-7136}\inst{\ref{aff75},\ref{aff13}}
\and L.~Moscardini\orcid{0000-0002-3473-6716}\inst{\ref{aff75},\ref{aff13},\ref{aff14}}
\and R.~Nakajima\orcid{0009-0009-1213-7040}\inst{\ref{aff73}}
\and C.~Neissner\orcid{0000-0001-8524-4968}\inst{\ref{aff80},\ref{aff29}}
\and R.~C.~Nichol\orcid{0000-0003-0939-6518}\inst{\ref{aff6}}
\and S.-M.~Niemi\inst{\ref{aff26}}
\and J.~W.~Nightingale\orcid{0000-0002-8987-7401}\inst{\ref{aff81}}
\and C.~Padilla\orcid{0000-0001-7951-0166}\inst{\ref{aff80}}
\and S.~Paltani\orcid{0000-0002-8108-9179}\inst{\ref{aff46}}
\and F.~Pasian\orcid{0000-0002-4869-3227}\inst{\ref{aff9}}
\and K.~Pedersen\inst{\ref{aff82}}
\and W.~J.~Percival\orcid{0000-0002-0644-5727}\inst{\ref{aff83},\ref{aff84},\ref{aff85}}
\and V.~Pettorino\inst{\ref{aff26}}
\and S.~Pires\orcid{0000-0002-0249-2104}\inst{\ref{aff86}}
\and G.~Polenta\orcid{0000-0003-4067-9196}\inst{\ref{aff16}}
\and M.~Poncet\inst{\ref{aff87}}
\and L.~A.~Popa\inst{\ref{aff88}}
\and L.~Pozzetti\orcid{0000-0001-7085-0412}\inst{\ref{aff13}}
\and F.~Raison\orcid{0000-0002-7819-6918}\inst{\ref{aff54}}
\and R.~Rebolo\inst{\ref{aff35},\ref{aff89},\ref{aff90}}
\and A.~Renzi\orcid{0000-0001-9856-1970}\inst{\ref{aff91},\ref{aff48}}
\and J.~Rhodes\orcid{0000-0002-4485-8549}\inst{\ref{aff58}}
\and G.~Riccio\inst{\ref{aff20}}
\and E.~Romelli\orcid{0000-0003-3069-9222}\inst{\ref{aff9}}
\and M.~Roncarelli\orcid{0000-0001-9587-7822}\inst{\ref{aff13}}
\and E.~Rossetti\orcid{0000-0003-0238-4047}\inst{\ref{aff12}}
\and R.~Saglia\orcid{0000-0003-0378-7032}\inst{\ref{aff53},\ref{aff54}}
\and Z.~Sakr\orcid{0000-0002-4823-3757}\inst{\ref{aff92},\ref{aff93},\ref{aff94}}
\and A.~G.~S\'anchez\orcid{0000-0003-1198-831X}\inst{\ref{aff54}}
\and D.~Sapone\orcid{0000-0001-7089-4503}\inst{\ref{aff95}}
\and B.~Sartoris\orcid{0000-0003-1337-5269}\inst{\ref{aff53},\ref{aff9}}
\and J.~A.~Schewtschenko\orcid{0000-0002-4913-6393}\inst{\ref{aff36}}
\and P.~Schneider\orcid{0000-0001-8561-2679}\inst{\ref{aff73}}
\and T.~Schrabback\orcid{0000-0002-6987-7834}\inst{\ref{aff96}}
\and A.~Secroun\orcid{0000-0003-0505-3710}\inst{\ref{aff49}}
\and E.~Sefusatti\orcid{0000-0003-0473-1567}\inst{\ref{aff9},\ref{aff8},\ref{aff10}}
\and G.~Seidel\orcid{0000-0003-2907-353X}\inst{\ref{aff1}}
\and M.~Seiffert\orcid{0000-0002-7536-9393}\inst{\ref{aff58}}
\and S.~Serrano\orcid{0000-0002-0211-2861}\inst{\ref{aff97},\ref{aff98},\ref{aff99}}
\and P.~Simon\inst{\ref{aff73}}
\and C.~Sirignano\orcid{0000-0002-0995-7146}\inst{\ref{aff91},\ref{aff48}}
\and G.~Sirri\orcid{0000-0003-2626-2853}\inst{\ref{aff14}}
\and L.~Stanco\orcid{0000-0002-9706-5104}\inst{\ref{aff48}}
\and J.~Steinwagner\orcid{0000-0001-7443-1047}\inst{\ref{aff54}}
\and P.~Tallada-Cresp\'{i}\orcid{0000-0002-1336-8328}\inst{\ref{aff28},\ref{aff29}}
\and D.~Tavagnacco\orcid{0000-0001-7475-9894}\inst{\ref{aff9}}
\and A.~N.~Taylor\inst{\ref{aff36}}
\and I.~Tereno\inst{\ref{aff44},\ref{aff100}}
\and S.~Toft\orcid{0000-0003-3631-7176}\inst{\ref{aff101},\ref{aff102}}
\and R.~Toledo-Moreo\orcid{0000-0002-2997-4859}\inst{\ref{aff103}}
\and F.~Torradeflot\orcid{0000-0003-1160-1517}\inst{\ref{aff29},\ref{aff28}}
\and I.~Tutusaus\orcid{0000-0002-3199-0399}\inst{\ref{aff93}}
\and L.~Valenziano\orcid{0000-0002-1170-0104}\inst{\ref{aff13},\ref{aff50}}
\and J.~Valiviita\orcid{0000-0001-6225-3693}\inst{\ref{aff66},\ref{aff67}}
\and T.~Vassallo\orcid{0000-0001-6512-6358}\inst{\ref{aff53},\ref{aff9}}
\and G.~Verdoes~Kleijn\orcid{0000-0001-5803-2580}\inst{\ref{aff104}}
\and A.~Veropalumbo\orcid{0000-0003-2387-1194}\inst{\ref{aff7},\ref{aff18},\ref{aff17}}
\and Y.~Wang\orcid{0000-0002-4749-2984}\inst{\ref{aff105}}
\and J.~Weller\orcid{0000-0002-8282-2010}\inst{\ref{aff53},\ref{aff54}}
\and A.~Zacchei\orcid{0000-0003-0396-1192}\inst{\ref{aff9},\ref{aff8}}
\and G.~Zamorani\orcid{0000-0002-2318-301X}\inst{\ref{aff13}}
\and I.~A.~Zinchenko\orcid{0000-0002-2944-2449}\inst{\ref{aff53}}
\and E.~Zucca\orcid{0000-0002-5845-8132}\inst{\ref{aff13}}
\and V.~Allevato\orcid{0000-0001-7232-5152}\inst{\ref{aff20}}
\and M.~Ballardini\orcid{0000-0003-4481-3559}\inst{\ref{aff106},\ref{aff107},\ref{aff13}}
\and M.~Bolzonella\orcid{0000-0003-3278-4607}\inst{\ref{aff13}}
\and E.~Bozzo\orcid{0000-0002-8201-1525}\inst{\ref{aff46}}
\and C.~Burigana\orcid{0000-0002-3005-5796}\inst{\ref{aff108},\ref{aff50}}
\and R.~Cabanac\orcid{0000-0001-6679-2600}\inst{\ref{aff93}}
\and M.~Calabrese\orcid{0000-0002-2637-2422}\inst{\ref{aff109},\ref{aff52}}
\and P.~Casenove\inst{\ref{aff87}}
\and D.~Di~Ferdinando\inst{\ref{aff14}}
\and J.~A.~Escartin~Vigo\inst{\ref{aff54}}
\and L.~Gabarra\orcid{0000-0002-8486-8856}\inst{\ref{aff110}}
\and S.~Matthew\orcid{0000-0001-8448-1697}\inst{\ref{aff36}}
\and N.~Mauri\orcid{0000-0001-8196-1548}\inst{\ref{aff34},\ref{aff14}}
\and R.~B.~Metcalf\orcid{0000-0003-3167-2574}\inst{\ref{aff75},\ref{aff13}}
\and A.~A.~Nucita\inst{\ref{aff111},\ref{aff112},\ref{aff113}}
\and A.~Pezzotta\orcid{0000-0003-0726-2268}\inst{\ref{aff54}}
\and M.~P\"ontinen\orcid{0000-0001-5442-2530}\inst{\ref{aff66}}
\and C.~Porciani\orcid{0000-0002-7797-2508}\inst{\ref{aff73}}
\and V.~Scottez\inst{\ref{aff114},\ref{aff115}}
\and M.~Tenti\orcid{0000-0002-4254-5901}\inst{\ref{aff14}}
\and M.~Viel\orcid{0000-0002-2642-5707}\inst{\ref{aff8},\ref{aff9},\ref{aff11},\ref{aff10},\ref{aff116}}
\and M.~Wiesmann\orcid{0009-0000-8199-5860}\inst{\ref{aff57}}
\and Y.~Akrami\orcid{0000-0002-2407-7956}\inst{\ref{aff117},\ref{aff118}}
\and I.~T.~Andika\orcid{0000-0001-6102-9526}\inst{\ref{aff119},\ref{aff120}}
\and S.~Anselmi\orcid{0000-0002-3579-9583}\inst{\ref{aff48},\ref{aff91},\ref{aff121}}
\and M.~Archidiacono\orcid{0000-0003-4952-9012}\inst{\ref{aff55},\ref{aff56}}
\and F.~Atrio-Barandela\orcid{0000-0002-2130-2513}\inst{\ref{aff122}}
\and D.~Bertacca\orcid{0000-0002-2490-7139}\inst{\ref{aff91},\ref{aff15},\ref{aff48}}
\and M.~Bethermin\orcid{0000-0002-3915-2015}\inst{\ref{aff123}}
\and A.~Blanchard\orcid{0000-0001-8555-9003}\inst{\ref{aff93}}
\and L.~Blot\orcid{0000-0002-9622-7167}\inst{\ref{aff124},\ref{aff121}}
\and S.~Borgani\orcid{0000-0001-6151-6439}\inst{\ref{aff125},\ref{aff8},\ref{aff9},\ref{aff10},\ref{aff116}}
\and M.~L.~Brown\orcid{0000-0002-0370-8077}\inst{\ref{aff37}}
\and S.~Bruton\orcid{0000-0002-6503-5218}\inst{\ref{aff126}}
\and A.~Calabro\orcid{0000-0003-2536-1614}\inst{\ref{aff31}}
\and A.~Cappi\inst{\ref{aff13},\ref{aff127}}
\and F.~Caro\inst{\ref{aff31}}
\and C.~S.~Carvalho\inst{\ref{aff100}}
\and T.~Castro\orcid{0000-0002-6292-3228}\inst{\ref{aff9},\ref{aff10},\ref{aff8},\ref{aff116}}
\and R.~Chary\orcid{0000-0001-7583-0621}\inst{\ref{aff105},\ref{aff128}}
\and F.~Cogato\orcid{0000-0003-4632-6113}\inst{\ref{aff75},\ref{aff13}}
\and S.~Conseil\orcid{0000-0002-3657-4191}\inst{\ref{aff39}}
\and A.~R.~Cooray\orcid{0000-0002-3892-0190}\inst{\ref{aff129}}
\and O.~Cucciati\orcid{0000-0002-9336-7551}\inst{\ref{aff13}}
\and S.~Davini\orcid{0000-0003-3269-1718}\inst{\ref{aff18}}
\and F.~De~Paolis\orcid{0000-0001-6460-7563}\inst{\ref{aff111},\ref{aff112},\ref{aff113}}
\and G.~Desprez\orcid{0000-0001-8325-1742}\inst{\ref{aff104}}
\and A.~D\'iaz-S\'anchez\orcid{0000-0003-0748-4768}\inst{\ref{aff130}}
\and S.~Di~Domizio\orcid{0000-0003-2863-5895}\inst{\ref{aff17},\ref{aff18}}
\and J.~M.~Diego\orcid{0000-0001-9065-3926}\inst{\ref{aff131}}
\and P.~Dimauro\orcid{0000-0001-7399-2854}\inst{\ref{aff31},\ref{aff132}}
\and A.~Enia\orcid{0000-0002-0200-2857}\inst{\ref{aff12},\ref{aff13}}
\and Y.~Fang\inst{\ref{aff53}}
\and A.~M.~N.~Ferguson\inst{\ref{aff36}}
\and A.~G.~Ferrari\orcid{0009-0005-5266-4110}\inst{\ref{aff14}}
\and A.~Finoguenov\orcid{0000-0002-4606-5403}\inst{\ref{aff66}}
\and A.~Franco\orcid{0000-0002-4761-366X}\inst{\ref{aff112},\ref{aff111},\ref{aff113}}
\and K.~Ganga\orcid{0000-0001-8159-8208}\inst{\ref{aff77}}
\and J.~Garc\'ia-Bellido\orcid{0000-0002-9370-8360}\inst{\ref{aff117}}
\and T.~Gasparetto\orcid{0000-0002-7913-4866}\inst{\ref{aff9}}
\and V.~Gautard\inst{\ref{aff2}}
\and E.~Gaztanaga\orcid{0000-0001-9632-0815}\inst{\ref{aff99},\ref{aff97},\ref{aff133}}
\and F.~Giacomini\orcid{0000-0002-3129-2814}\inst{\ref{aff14}}
\and F.~Gianotti\orcid{0000-0003-4666-119X}\inst{\ref{aff13}}
\and G.~Gozaliasl\orcid{0000-0002-0236-919X}\inst{\ref{aff134},\ref{aff66}}
\and A.~Gregorio\orcid{0000-0003-4028-8785}\inst{\ref{aff125},\ref{aff9},\ref{aff10}}
\and M.~Guidi\orcid{0000-0001-9408-1101}\inst{\ref{aff12},\ref{aff13}}
\and C.~M.~Gutierrez\orcid{0000-0001-7854-783X}\inst{\ref{aff135}}
\and A.~Hall\orcid{0000-0002-3139-8651}\inst{\ref{aff36}}
\and W.~G.~Hartley\inst{\ref{aff46}}
\and S.~Hemmati\orcid{0000-0003-2226-5395}\inst{\ref{aff136}}
\and C.~Hern\'andez-Monteagudo\orcid{0000-0001-5471-9166}\inst{\ref{aff90},\ref{aff35}}
\and H.~Hildebrandt\orcid{0000-0002-9814-3338}\inst{\ref{aff137}}
\and J.~Hjorth\orcid{0000-0002-4571-2306}\inst{\ref{aff82}}
\and J.~J.~E.~Kajava\orcid{0000-0002-3010-8333}\inst{\ref{aff138},\ref{aff139}}
\and Y.~Kang\orcid{0009-0000-8588-7250}\inst{\ref{aff46}}
\and V.~Kansal\orcid{0000-0002-4008-6078}\inst{\ref{aff140},\ref{aff141}}
\and D.~Karagiannis\orcid{0000-0002-4927-0816}\inst{\ref{aff106},\ref{aff142}}
\and K.~Kiiveri\inst{\ref{aff64}}
\and C.~C.~Kirkpatrick\inst{\ref{aff64}}
\and S.~Kruk\orcid{0000-0001-8010-8879}\inst{\ref{aff5}}
\and J.~Le~Graet\orcid{0000-0001-6523-7971}\inst{\ref{aff49}}
\and L.~Legrand\orcid{0000-0003-0610-5252}\inst{\ref{aff143},\ref{aff144}}
\and M.~Lembo\orcid{0000-0002-5271-5070}\inst{\ref{aff106},\ref{aff107}}
\and F.~Lepori\orcid{0009-0000-5061-7138}\inst{\ref{aff145}}
\and G.~Leroy\orcid{0009-0004-2523-4425}\inst{\ref{aff146},\ref{aff76}}
\and J.~Lesgourgues\orcid{0000-0001-7627-353X}\inst{\ref{aff30}}
\and L.~Leuzzi\orcid{0009-0006-4479-7017}\inst{\ref{aff75},\ref{aff13}}
\and T.~I.~Liaudat\orcid{0000-0002-9104-314X}\inst{\ref{aff147}}
\and S.~J.~Liu\orcid{0000-0001-7680-2139}\inst{\ref{aff47}}
\and A.~Loureiro\orcid{0000-0002-4371-0876}\inst{\ref{aff148},\ref{aff149}}
\and J.~Macias-Perez\orcid{0000-0002-5385-2763}\inst{\ref{aff150}}
\and G.~Maggio\orcid{0000-0003-4020-4836}\inst{\ref{aff9}}
\and M.~Magliocchetti\orcid{0000-0001-9158-4838}\inst{\ref{aff47}}
\and F.~Mannucci\orcid{0000-0002-4803-2381}\inst{\ref{aff151}}
\and R.~Maoli\orcid{0000-0002-6065-3025}\inst{\ref{aff152},\ref{aff31}}
\and J.~Mart\'{i}n-Fleitas\orcid{0000-0002-8594-569X}\inst{\ref{aff79}}
\and C.~J.~A.~P.~Martins\orcid{0000-0002-4886-9261}\inst{\ref{aff153},\ref{aff21}}
\and L.~Maurin\orcid{0000-0002-8406-0857}\inst{\ref{aff4}}
\and M.~Miluzio\inst{\ref{aff5},\ref{aff154}}
\and P.~Monaco\orcid{0000-0003-2083-7564}\inst{\ref{aff125},\ref{aff9},\ref{aff10},\ref{aff8}}
\and A.~Montoro\orcid{0000-0003-4730-8590}\inst{\ref{aff99},\ref{aff97}}
\and C.~Moretti\orcid{0000-0003-3314-8936}\inst{\ref{aff11},\ref{aff116},\ref{aff9},\ref{aff8},\ref{aff10}}
\and G.~Morgante\inst{\ref{aff13}}
\and S.~Nadathur\orcid{0000-0001-9070-3102}\inst{\ref{aff133}}
\and K.~Naidoo\orcid{0000-0002-9182-1802}\inst{\ref{aff133}}
\and P.~Natoli\orcid{0000-0003-0126-9100}\inst{\ref{aff106},\ref{aff107}}
\and A.~Navarro-Alsina\orcid{0000-0002-3173-2592}\inst{\ref{aff73}}
\and S.~Nesseris\orcid{0000-0002-0567-0324}\inst{\ref{aff117}}
\and F.~Passalacqua\orcid{0000-0002-8606-4093}\inst{\ref{aff91},\ref{aff48}}
\and L.~Patrizii\inst{\ref{aff14}}
\and A.~Pisani\orcid{0000-0002-6146-4437}\inst{\ref{aff49},\ref{aff155}}
\and D.~Potter\orcid{0000-0002-0757-5195}\inst{\ref{aff145}}
\and S.~Quai\orcid{0000-0002-0449-8163}\inst{\ref{aff75},\ref{aff13}}
\and M.~Radovich\orcid{0000-0002-3585-866X}\inst{\ref{aff15}}
\and P.~Reimberg\orcid{0000-0003-3410-0280}\inst{\ref{aff114}}
\and I.~Risso\orcid{0000-0003-2525-7761}\inst{\ref{aff156}}
\and S.~Sacquegna\orcid{0000-0002-8433-6630}\inst{\ref{aff111},\ref{aff112},\ref{aff113}}
\and M.~Sahl\'en\orcid{0000-0003-0973-4804}\inst{\ref{aff157}}
\and E.~Sarpa\orcid{0000-0002-1256-655X}\inst{\ref{aff11},\ref{aff116},\ref{aff10}}
\and A.~Schneider\orcid{0000-0001-7055-8104}\inst{\ref{aff145}}
\and M.~Schultheis\inst{\ref{aff127}}
\and D.~Sciotti\orcid{0009-0008-4519-2620}\inst{\ref{aff31},\ref{aff74}}
\and E.~Sellentin\inst{\ref{aff158},\ref{aff3}}
\and M.~Sereno\orcid{0000-0003-0302-0325}\inst{\ref{aff13},\ref{aff14}}
\and A.~Shulevski\orcid{0000-0002-1827-0469}\inst{\ref{aff159},\ref{aff104},\ref{aff160},\ref{aff161}}
\and L.~C.~Smith\orcid{0000-0002-3259-2771}\inst{\ref{aff162}}
\and J.~Stadel\orcid{0000-0001-7565-8622}\inst{\ref{aff145}}
\and K.~Tanidis\orcid{0000-0001-9843-5130}\inst{\ref{aff110}}
\and C.~Tao\orcid{0000-0001-7961-8177}\inst{\ref{aff49}}
\and G.~Testera\inst{\ref{aff18}}
\and R.~Teyssier\orcid{0000-0001-7689-0933}\inst{\ref{aff155}}
\and S.~Tosi\orcid{0000-0002-7275-9193}\inst{\ref{aff17},\ref{aff156}}
\and A.~Troja\orcid{0000-0003-0239-4595}\inst{\ref{aff91},\ref{aff48}}
\and M.~Tucci\inst{\ref{aff46}}
\and C.~Valieri\inst{\ref{aff14}}
\and A.~Venhola\orcid{0000-0001-6071-4564}\inst{\ref{aff163}}
\and D.~Vergani\orcid{0000-0003-0898-2216}\inst{\ref{aff13}}
\and G.~Verza\orcid{0000-0002-1886-8348}\inst{\ref{aff164}}
\and P.~Vielzeuf\orcid{0000-0003-2035-9339}\inst{\ref{aff49}}
\and N.~A.~Walton\orcid{0000-0003-3983-8778}\inst{\ref{aff162}}}
										   
\institute{Max-Planck-Institut f\"ur Astronomie, K\"onigstuhl 17, 69117 Heidelberg, Germany\label{aff1}
\and
CEA Saclay, DFR/IRFU, Service d'Astrophysique, Bat. 709, 91191 Gif-sur-Yvette, France\label{aff2}
\and
Leiden Observatory, Leiden University, Einsteinweg 55, 2333 CC Leiden, The Netherlands\label{aff3}
\and
Universit\'e Paris-Saclay, CNRS, Institut d'astrophysique spatiale, 91405, Orsay, France\label{aff4}
\and
ESAC/ESA, Camino Bajo del Castillo, s/n., Urb. Villafranca del Castillo, 28692 Villanueva de la Ca\~nada, Madrid, Spain\label{aff5}
\and
School of Mathematics and Physics, University of Surrey, Guildford, Surrey, GU2 7XH, UK\label{aff6}
\and
INAF-Osservatorio Astronomico di Brera, Via Brera 28, 20122 Milano, Italy\label{aff7}
\and
IFPU, Institute for Fundamental Physics of the Universe, via Beirut 2, 34151 Trieste, Italy\label{aff8}
\and
INAF-Osservatorio Astronomico di Trieste, Via G. B. Tiepolo 11, 34143 Trieste, Italy\label{aff9}
\and
INFN, Sezione di Trieste, Via Valerio 2, 34127 Trieste TS, Italy\label{aff10}
\and
SISSA, International School for Advanced Studies, Via Bonomea 265, 34136 Trieste TS, Italy\label{aff11}
\and
Dipartimento di Fisica e Astronomia, Universit\`a di Bologna, Via Gobetti 93/2, 40129 Bologna, Italy\label{aff12}
\and
INAF-Osservatorio di Astrofisica e Scienza dello Spazio di Bologna, Via Piero Gobetti 93/3, 40129 Bologna, Italy\label{aff13}
\and
INFN-Sezione di Bologna, Viale Berti Pichat 6/2, 40127 Bologna, Italy\label{aff14}
\and
INAF-Osservatorio Astronomico di Padova, Via dell'Osservatorio 5, 35122 Padova, Italy\label{aff15}
\and
Space Science Data Center, Italian Space Agency, via del Politecnico snc, 00133 Roma, Italy\label{aff16}
\and
Dipartimento di Fisica, Universit\`a di Genova, Via Dodecaneso 33, 16146, Genova, Italy\label{aff17}
\and
INFN-Sezione di Genova, Via Dodecaneso 33, 16146, Genova, Italy\label{aff18}
\and
Department of Physics "E. Pancini", University Federico II, Via Cinthia 6, 80126, Napoli, Italy\label{aff19}
\and
INAF-Osservatorio Astronomico di Capodimonte, Via Moiariello 16, 80131 Napoli, Italy\label{aff20}
\and
Instituto de Astrof\'isica e Ci\^encias do Espa\c{c}o, Universidade do Porto, CAUP, Rua das Estrelas, PT4150-762 Porto, Portugal\label{aff21}
\and
Faculdade de Ci\^encias da Universidade do Porto, Rua do Campo de Alegre, 4150-007 Porto, Portugal\label{aff22}
\and
Dipartimento di Fisica, Universit\`a degli Studi di Torino, Via P. Giuria 1, 10125 Torino, Italy\label{aff23}
\and
INFN-Sezione di Torino, Via P. Giuria 1, 10125 Torino, Italy\label{aff24}
\and
INAF-Osservatorio Astrofisico di Torino, Via Osservatorio 20, 10025 Pino Torinese (TO), Italy\label{aff25}
\and
European Space Agency/ESTEC, Keplerlaan 1, 2201 AZ Noordwijk, The Netherlands\label{aff26}
\and
Institute Lorentz, Leiden University, Niels Bohrweg 2, 2333 CA Leiden, The Netherlands\label{aff27}
\and
Centro de Investigaciones Energ\'eticas, Medioambientales y Tecnol\'ogicas (CIEMAT), Avenida Complutense 40, 28040 Madrid, Spain\label{aff28}
\and
Port d'Informaci\'{o} Cient\'{i}fica, Campus UAB, C. Albareda s/n, 08193 Bellaterra (Barcelona), Spain\label{aff29}
\and
Institute for Theoretical Particle Physics and Cosmology (TTK), RWTH Aachen University, 52056 Aachen, Germany\label{aff30}
\and
INAF-Osservatorio Astronomico di Roma, Via Frascati 33, 00078 Monteporzio Catone, Italy\label{aff31}
\and
INFN section of Naples, Via Cinthia 6, 80126, Napoli, Italy\label{aff32}
\and
Institute for Astronomy, University of Hawaii, 2680 Woodlawn Drive, Honolulu, HI 96822, USA\label{aff33}
\and
Dipartimento di Fisica e Astronomia "Augusto Righi" - Alma Mater Studiorum Universit\`a di Bologna, Viale Berti Pichat 6/2, 40127 Bologna, Italy\label{aff34}
\and
Instituto de Astrof\'{\i}sica de Canarias, V\'{\i}a L\'actea, 38205 La Laguna, Tenerife, Spain\label{aff35}
\and
Institute for Astronomy, University of Edinburgh, Royal Observatory, Blackford Hill, Edinburgh EH9 3HJ, UK\label{aff36}
\and
Jodrell Bank Centre for Astrophysics, Department of Physics and Astronomy, University of Manchester, Oxford Road, Manchester M13 9PL, UK\label{aff37}
\and
European Space Agency/ESRIN, Largo Galileo Galilei 1, 00044 Frascati, Roma, Italy\label{aff38}
\and
Universit\'e Claude Bernard Lyon 1, CNRS/IN2P3, IP2I Lyon, UMR 5822, Villeurbanne, F-69100, France\label{aff39}
\and
Institute of Physics, Laboratory of Astrophysics, Ecole Polytechnique F\'ed\'erale de Lausanne (EPFL), Observatoire de Sauverny, 1290 Versoix, Switzerland\label{aff40}
\and
Institut de Ci\`{e}ncies del Cosmos (ICCUB), Universitat de Barcelona (IEEC-UB), Mart\'{i} i Franqu\`{e}s 1, 08028 Barcelona, Spain\label{aff41}
\and
Instituci\'o Catalana de Recerca i Estudis Avan\c{c}ats (ICREA), Passeig de Llu\'{\i}s Companys 23, 08010 Barcelona, Spain\label{aff42}
\and
UCB Lyon 1, CNRS/IN2P3, IUF, IP2I Lyon, 4 rue Enrico Fermi, 69622 Villeurbanne, France\label{aff43}
\and
Departamento de F\'isica, Faculdade de Ci\^encias, Universidade de Lisboa, Edif\'icio C8, Campo Grande, PT1749-016 Lisboa, Portugal\label{aff44}
\and
Instituto de Astrof\'isica e Ci\^encias do Espa\c{c}o, Faculdade de Ci\^encias, Universidade de Lisboa, Campo Grande, 1749-016 Lisboa, Portugal\label{aff45}
\and
Department of Astronomy, University of Geneva, ch. d'Ecogia 16, 1290 Versoix, Switzerland\label{aff46}
\and
INAF-Istituto di Astrofisica e Planetologia Spaziali, via del Fosso del Cavaliere, 100, 00100 Roma, Italy\label{aff47}
\and
INFN-Padova, Via Marzolo 8, 35131 Padova, Italy\label{aff48}
\and
Aix-Marseille Universit\'e, CNRS/IN2P3, CPPM, Marseille, France\label{aff49}
\and
INFN-Bologna, Via Irnerio 46, 40126 Bologna, Italy\label{aff50}
\and
School of Physics, HH Wills Physics Laboratory, University of Bristol, Tyndall Avenue, Bristol, BS8 1TL, UK\label{aff51}
\and
INAF-IASF Milano, Via Alfonso Corti 12, 20133 Milano, Italy\label{aff52}
\and
Universit\"ats-Sternwarte M\"unchen, Fakult\"at f\"ur Physik, Ludwig-Maximilians-Universit\"at M\"unchen, Scheinerstrasse 1, 81679 M\"unchen, Germany\label{aff53}
\and
Max Planck Institute for Extraterrestrial Physics, Giessenbachstr. 1, 85748 Garching, Germany\label{aff54}
\and
Dipartimento di Fisica "Aldo Pontremoli", Universit\`a degli Studi di Milano, Via Celoria 16, 20133 Milano, Italy\label{aff55}
\and
INFN-Sezione di Milano, Via Celoria 16, 20133 Milano, Italy\label{aff56}
\and
Institute of Theoretical Astrophysics, University of Oslo, P.O. Box 1029 Blindern, 0315 Oslo, Norway\label{aff57}
\and
Jet Propulsion Laboratory, California Institute of Technology, 4800 Oak Grove Drive, Pasadena, CA, 91109, USA\label{aff58}
\and
Felix Hormuth Engineering, Goethestr. 17, 69181 Leimen, Germany\label{aff59}
\and
Technical University of Denmark, Elektrovej 327, 2800 Kgs. Lyngby, Denmark\label{aff60}
\and
Cosmic Dawn Center (DAWN), Denmark\label{aff61}
\and
Institut d'Astrophysique de Paris, UMR 7095, CNRS, and Sorbonne Universit\'e, 98 bis boulevard Arago, 75014 Paris, France\label{aff62}
\and
NASA Goddard Space Flight Center, Greenbelt, MD 20771, USA\label{aff63}
\and
Department of Physics and Helsinki Institute of Physics, Gustaf H\"allstr\"omin katu 2, 00014 University of Helsinki, Finland\label{aff64}
\and
Universit\'e de Gen\`eve, D\'epartement de Physique Th\'eorique and Centre for Astroparticle Physics, 24 quai Ernest-Ansermet, CH-1211 Gen\`eve 4, Switzerland\label{aff65}
\and
Department of Physics, P.O. Box 64, 00014 University of Helsinki, Finland\label{aff66}
\and
Helsinki Institute of Physics, Gustaf H{\"a}llstr{\"o}min katu 2, University of Helsinki, Helsinki, Finland\label{aff67}
\and
Laboratoire d'etude de l'Univers et des phenomenes eXtremes, Observatoire de Paris, Universit\'e PSL, Sorbonne Universit\'e, CNRS, 92190 Meudon, France\label{aff68}
\and
Aix-Marseille Universit\'e, CNRS, CNES, LAM, Marseille, France\label{aff69}
\and
SKA Observatory, Jodrell Bank, Lower Withington, Macclesfield, Cheshire SK11 9FT, UK\label{aff70}
\and
Centre de Calcul de l'IN2P3/CNRS, 21 avenue Pierre de Coubertin 69627 Villeurbanne Cedex, France\label{aff71}
\and
University of Applied Sciences and Arts of Northwestern Switzerland, School of Engineering, 5210 Windisch, Switzerland\label{aff72}
\and
Universit\"at Bonn, Argelander-Institut f\"ur Astronomie, Auf dem H\"ugel 71, 53121 Bonn, Germany\label{aff73}
\and
INFN-Sezione di Roma, Piazzale Aldo Moro, 2 - c/o Dipartimento di Fisica, Edificio G. Marconi, 00185 Roma, Italy\label{aff74}
\and
Dipartimento di Fisica e Astronomia "Augusto Righi" - Alma Mater Studiorum Universit\`a di Bologna, via Piero Gobetti 93/2, 40129 Bologna, Italy\label{aff75}
\and
Department of Physics, Institute for Computational Cosmology, Durham University, South Road, Durham, DH1 3LE, UK\label{aff76}
\and
Universit\'e Paris Cit\'e, CNRS, Astroparticule et Cosmologie, 75013 Paris, France\label{aff77}
\and
CNRS-UCB International Research Laboratory, Centre Pierre Binetruy, IRL2007, CPB-IN2P3, Berkeley, USA\label{aff78}
\and
Aurora Technology for European Space Agency (ESA), Camino bajo del Castillo, s/n, Urbanizacion Villafranca del Castillo, Villanueva de la Ca\~nada, 28692 Madrid, Spain\label{aff79}
\and
Institut de F\'{i}sica d'Altes Energies (IFAE), The Barcelona Institute of Science and Technology, Campus UAB, 08193 Bellaterra (Barcelona), Spain\label{aff80}
\and
School of Mathematics, Statistics and Physics, Newcastle University, Herschel Building, Newcastle-upon-Tyne, NE1 7RU, UK\label{aff81}
\and
DARK, Niels Bohr Institute, University of Copenhagen, Jagtvej 155, 2200 Copenhagen, Denmark\label{aff82}
\and
Waterloo Centre for Astrophysics, University of Waterloo, Waterloo, Ontario N2L 3G1, Canada\label{aff83}
\and
Department of Physics and Astronomy, University of Waterloo, Waterloo, Ontario N2L 3G1, Canada\label{aff84}
\and
Perimeter Institute for Theoretical Physics, Waterloo, Ontario N2L 2Y5, Canada\label{aff85}
\and
Universit\'e Paris-Saclay, Universit\'e Paris Cit\'e, CEA, CNRS, AIM, 91191, Gif-sur-Yvette, France\label{aff86}
\and
Centre National d'Etudes Spatiales -- Centre spatial de Toulouse, 18 avenue Edouard Belin, 31401 Toulouse Cedex 9, France\label{aff87}
\and
Institute of Space Science, Str. Atomistilor, nr. 409 M\u{a}gurele, Ilfov, 077125, Romania\label{aff88}
\and
Consejo Superior de Investigaciones Cientificas, Calle Serrano 117, 28006 Madrid, Spain\label{aff89}
\and
Universidad de La Laguna, Departamento de Astrof\'{\i}sica, 38206 La Laguna, Tenerife, Spain\label{aff90}
\and
Dipartimento di Fisica e Astronomia "G. Galilei", Universit\`a di Padova, Via Marzolo 8, 35131 Padova, Italy\label{aff91}
\and
Institut f\"ur Theoretische Physik, University of Heidelberg, Philosophenweg 16, 69120 Heidelberg, Germany\label{aff92}
\and
Institut de Recherche en Astrophysique et Plan\'etologie (IRAP), Universit\'e de Toulouse, CNRS, UPS, CNES, 14 Av. Edouard Belin, 31400 Toulouse, France\label{aff93}
\and
Universit\'e St Joseph; Faculty of Sciences, Beirut, Lebanon\label{aff94}
\and
Departamento de F\'isica, FCFM, Universidad de Chile, Blanco Encalada 2008, Santiago, Chile\label{aff95}
\and
Universit\"at Innsbruck, Institut f\"ur Astro- und Teilchenphysik, Technikerstr. 25/8, 6020 Innsbruck, Austria\label{aff96}
\and
Institut d'Estudis Espacials de Catalunya (IEEC),  Edifici RDIT, Campus UPC, 08860 Castelldefels, Barcelona, Spain\label{aff97}
\and
Satlantis, University Science Park, Sede Bld 48940, Leioa-Bilbao, Spain\label{aff98}
\and
Institute of Space Sciences (ICE, CSIC), Campus UAB, Carrer de Can Magrans, s/n, 08193 Barcelona, Spain\label{aff99}
\and
Instituto de Astrof\'isica e Ci\^encias do Espa\c{c}o, Faculdade de Ci\^encias, Universidade de Lisboa, Tapada da Ajuda, 1349-018 Lisboa, Portugal\label{aff100}
\and
Cosmic Dawn Center (DAWN)\label{aff101}
\and
Niels Bohr Institute, University of Copenhagen, Jagtvej 128, 2200 Copenhagen, Denmark\label{aff102}
\and
Universidad Polit\'ecnica de Cartagena, Departamento de Electr\'onica y Tecnolog\'ia de Computadoras,  Plaza del Hospital 1, 30202 Cartagena, Spain\label{aff103}
\and
Kapteyn Astronomical Institute, University of Groningen, PO Box 800, 9700 AV Groningen, The Netherlands\label{aff104}
\and
Infrared Processing and Analysis Center, California Institute of Technology, Pasadena, CA 91125, USA\label{aff105}
\and
Dipartimento di Fisica e Scienze della Terra, Universit\`a degli Studi di Ferrara, Via Giuseppe Saragat 1, 44122 Ferrara, Italy\label{aff106}
\and
Istituto Nazionale di Fisica Nucleare, Sezione di Ferrara, Via Giuseppe Saragat 1, 44122 Ferrara, Italy\label{aff107}
\and
INAF, Istituto di Radioastronomia, Via Piero Gobetti 101, 40129 Bologna, Italy\label{aff108}
\and
Astronomical Observatory of the Autonomous Region of the Aosta Valley (OAVdA), Loc. Lignan 39, I-11020, Nus (Aosta Valley), Italy\label{aff109}
\and
Department of Physics, Oxford University, Keble Road, Oxford OX1 3RH, UK\label{aff110}
\and
Department of Mathematics and Physics E. De Giorgi, University of Salento, Via per Arnesano, CP-I93, 73100, Lecce, Italy\label{aff111}
\and
INFN, Sezione di Lecce, Via per Arnesano, CP-193, 73100, Lecce, Italy\label{aff112}
\and
INAF-Sezione di Lecce, c/o Dipartimento Matematica e Fisica, Via per Arnesano, 73100, Lecce, Italy\label{aff113}
\and
Institut d'Astrophysique de Paris, 98bis Boulevard Arago, 75014, Paris, France\label{aff114}
\and
ICL, Junia, Universit\'e Catholique de Lille, LITL, 59000 Lille, France\label{aff115}
\and
ICSC - Centro Nazionale di Ricerca in High Performance Computing, Big Data e Quantum Computing, Via Magnanelli 2, Bologna, Italy\label{aff116}
\and
Instituto de F\'isica Te\'orica UAM-CSIC, Campus de Cantoblanco, 28049 Madrid, Spain\label{aff117}
\and
CERCA/ISO, Department of Physics, Case Western Reserve University, 10900 Euclid Avenue, Cleveland, OH 44106, USA\label{aff118}
\and
Technical University of Munich, TUM School of Natural Sciences, Physics Department, James-Franck-Str.~1, 85748 Garching, Germany\label{aff119}
\and
Max-Planck-Institut f\"ur Astrophysik, Karl-Schwarzschild-Str.~1, 85748 Garching, Germany\label{aff120}
\and
Laboratoire Univers et Th\'eorie, Observatoire de Paris, Universit\'e PSL, Universit\'e Paris Cit\'e, CNRS, 92190 Meudon, France\label{aff121}
\and
Departamento de F{\'\i}sica Fundamental. Universidad de Salamanca. Plaza de la Merced s/n. 37008 Salamanca, Spain\label{aff122}
\and
Universit\'e de Strasbourg, CNRS, Observatoire astronomique de Strasbourg, UMR 7550, 67000 Strasbourg, France\label{aff123}
\and
Center for Data-Driven Discovery, Kavli IPMU (WPI), UTIAS, The University of Tokyo, Kashiwa, Chiba 277-8583, Japan\label{aff124}
\and
Dipartimento di Fisica - Sezione di Astronomia, Universit\`a di Trieste, Via Tiepolo 11, 34131 Trieste, Italy\label{aff125}
\and
California Institute of Technology, 1200 E California Blvd, Pasadena, CA 91125, USA\label{aff126}
\and
Universit\'e C\^{o}te d'Azur, Observatoire de la C\^{o}te d'Azur, CNRS, Laboratoire Lagrange, Bd de l'Observatoire, CS 34229, 06304 Nice cedex 4, France\label{aff127}
\and
University of California, Los Angeles, CA 90095-1562, USA\label{aff128}
\and
Department of Physics \& Astronomy, University of California Irvine, Irvine CA 92697, USA\label{aff129}
\and
Departamento F\'isica Aplicada, Universidad Polit\'ecnica de Cartagena, Campus Muralla del Mar, 30202 Cartagena, Murcia, Spain\label{aff130}
\and
Instituto de F\'isica de Cantabria, Edificio Juan Jord\'a, Avenida de los Castros, 39005 Santander, Spain\label{aff131}
\and
Observatorio Nacional, Rua General Jose Cristino, 77-Bairro Imperial de Sao Cristovao, Rio de Janeiro, 20921-400, Brazil\label{aff132}
\and
Institute of Cosmology and Gravitation, University of Portsmouth, Portsmouth PO1 3FX, UK\label{aff133}
\and
Department of Computer Science, Aalto University, PO Box 15400, Espoo, FI-00 076, Finland\label{aff134}
\and
Instituto de Astrof\'\i sica de Canarias, c/ Via Lactea s/n, La Laguna 38200, Spain. Departamento de Astrof\'\i sica de la Universidad de La Laguna, Avda. Francisco Sanchez, La Laguna, 38200, Spain\label{aff135}
\and
Caltech/IPAC, 1200 E. California Blvd., Pasadena, CA 91125, USA\label{aff136}
\and
Ruhr University Bochum, Faculty of Physics and Astronomy, Astronomical Institute (AIRUB), German Centre for Cosmological Lensing (GCCL), 44780 Bochum, Germany\label{aff137}
\and
Department of Physics and Astronomy, Vesilinnantie 5, 20014 University of Turku, Finland\label{aff138}
\and
Serco for European Space Agency (ESA), Camino bajo del Castillo, s/n, Urbanizacion Villafranca del Castillo, Villanueva de la Ca\~nada, 28692 Madrid, Spain\label{aff139}
\and
ARC Centre of Excellence for Dark Matter Particle Physics, Melbourne, Australia\label{aff140}
\and
Centre for Astrophysics \& Supercomputing, Swinburne University of Technology,  Hawthorn, Victoria 3122, Australia\label{aff141}
\and
Department of Physics and Astronomy, University of the Western Cape, Bellville, Cape Town, 7535, South Africa\label{aff142}
\and
DAMTP, Centre for Mathematical Sciences, Wilberforce Road, Cambridge CB3 0WA, UK\label{aff143}
\and
Kavli Institute for Cosmology Cambridge, Madingley Road, Cambridge, CB3 0HA, UK\label{aff144}
\and
Department of Astrophysics, University of Zurich, Winterthurerstrasse 190, 8057 Zurich, Switzerland\label{aff145}
\and
Department of Physics, Centre for Extragalactic Astronomy, Durham University, South Road, Durham, DH1 3LE, UK\label{aff146}
\and
IRFU, CEA, Universit\'e Paris-Saclay 91191 Gif-sur-Yvette Cedex, France\label{aff147}
\and
Oskar Klein Centre for Cosmoparticle Physics, Department of Physics, Stockholm University, Stockholm, SE-106 91, Sweden\label{aff148}
\and
Astrophysics Group, Blackett Laboratory, Imperial College London, London SW7 2AZ, UK\label{aff149}
\and
Univ. Grenoble Alpes, CNRS, Grenoble INP, LPSC-IN2P3, 53, Avenue des Martyrs, 38000, Grenoble, France\label{aff150}
\and
INAF-Osservatorio Astrofisico di Arcetri, Largo E. Fermi 5, 50125, Firenze, Italy\label{aff151}
\and
Dipartimento di Fisica, Sapienza Universit\`a di Roma, Piazzale Aldo Moro 2, 00185 Roma, Italy\label{aff152}
\and
Centro de Astrof\'{\i}sica da Universidade do Porto, Rua das Estrelas, 4150-762 Porto, Portugal\label{aff153}
\and
HE Space for European Space Agency (ESA), Camino bajo del Castillo, s/n, Urbanizacion Villafranca del Castillo, Villanueva de la Ca\~nada, 28692 Madrid, Spain\label{aff154}
\and
Department of Astrophysical Sciences, Peyton Hall, Princeton University, Princeton, NJ 08544, USA\label{aff155}
\and
INAF-Osservatorio Astronomico di Brera, Via Brera 28, 20122 Milano, Italy, and INFN-Sezione di Genova, Via Dodecaneso 33, 16146, Genova, Italy\label{aff156}
\and
Theoretical astrophysics, Department of Physics and Astronomy, Uppsala University, Box 515, 751 20 Uppsala, Sweden\label{aff157}
\and
Mathematical Institute, University of Leiden, Einsteinweg 55, 2333 CA Leiden, The Netherlands\label{aff158}
\and
ASTRON, the Netherlands Institute for Radio Astronomy, Postbus 2, 7990 AA, Dwingeloo, The Netherlands\label{aff159}
\and
Anton Pannekoek Institute for Astronomy, University of Amsterdam, Postbus 94249, 1090 GE Amsterdam, The Netherlands\label{aff160}
\and
Center for Advanced Interdisciplinary Research, Ss. Cyril and Methodius University in Skopje, Macedonia\label{aff161}
\and
Institute of Astronomy, University of Cambridge, Madingley Road, Cambridge CB3 0HA, UK\label{aff162}
\and
Space physics and astronomy research unit, University of Oulu, Pentti Kaiteran katu 1, FI-90014 Oulu, Finland\label{aff163}
\and
Center for Computational Astrophysics, Flatiron Institute, 162 5th Avenue, 10010, New York, NY, USA\label{aff164}}    

%
%
   \abstract{The Near-Infrared Spectrometer and Photometer (NISP) onboard \Euclid includes several optical elements in its path, which introduce artefacts into the data from non-nominal light paths. To ensure uncontaminated source photometry, these artefacts must be accurately accounted for. This paper focuses on two specific optical features in NISP's photometric data (NISP-P): ghosts caused by the telescope's dichroic beamsplitter, and the bandpass filters within the NISP fore-optics. Both ghost types  exhibit a characteristic morphology and are offset from the originating stars. The offsets are well modelled using 2D polynomials, with only stars brighter than approximately 10 magnitudes in each filter producing significant ghost contributions. The masking radii for these ghosts depend on both the source-star brightness and the filter wavelength, ranging from 20 to 40 pixels. We present the final relations and models used in the near-infrared (NIR) data pipeline to mask these ghosts for \Euclid's Quick Data Release (Q1).}
%
%
%
%
   \titlerunning{\Euclid: NISP-P optical ghosts}
   \authorrunning{Euclid Collaboration: K.\ Paterson et al.}
   
   \maketitle
%
%
%
%


\section{Introduction}
\label{sc:Intro}

\Euclid was launched in July 2023 and started its nominal observations, the Euclid Wide Survey (EWS), in February 2024. An overview of the mission, including early results from the Performance Verification (PV), is given in \citet{EuclidSkyOverview}. \Euclid's two instruments, VIS and the Near-Infrared Spectrometer and Photometer (NISP), are described in detail respectively in \citet{EuclidSkyVIS} and \citet{EuclidSkyNISP}. Both instruments can observe the sky simultaneously by means of a dichroic beamsplitter. The VIS optical path is purely reflective, whereas NISP implements the dichroic in transmission and carries three filters and four lenses in its fore-optics. For \Euclid to achieve its demanding scientific goals, accurate masking of unwanted optical features is essential for uncontaminated source photometry.

\begin{figure*}[t]
\centering
\includegraphics[width=0.9\hsize]{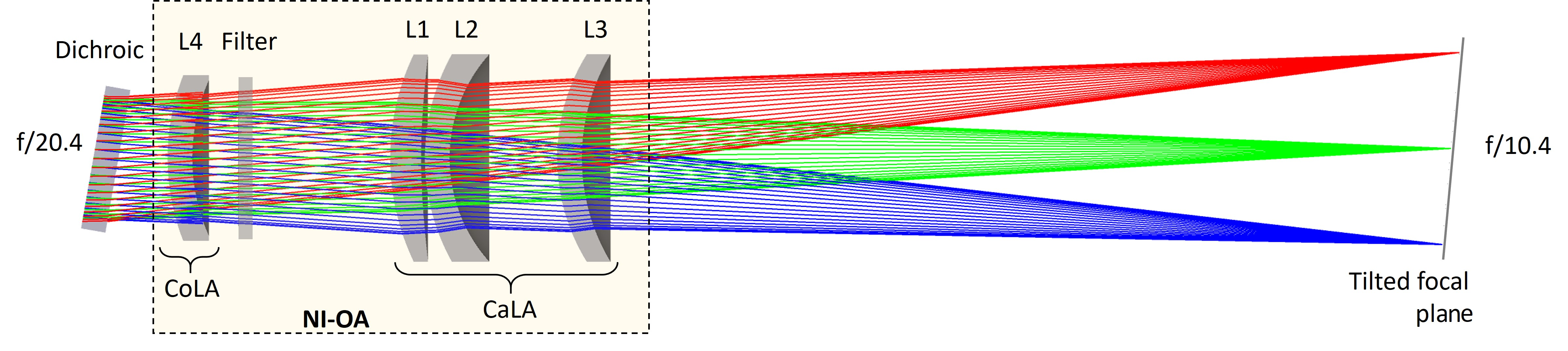}
\includegraphics[width=0.85\hsize]{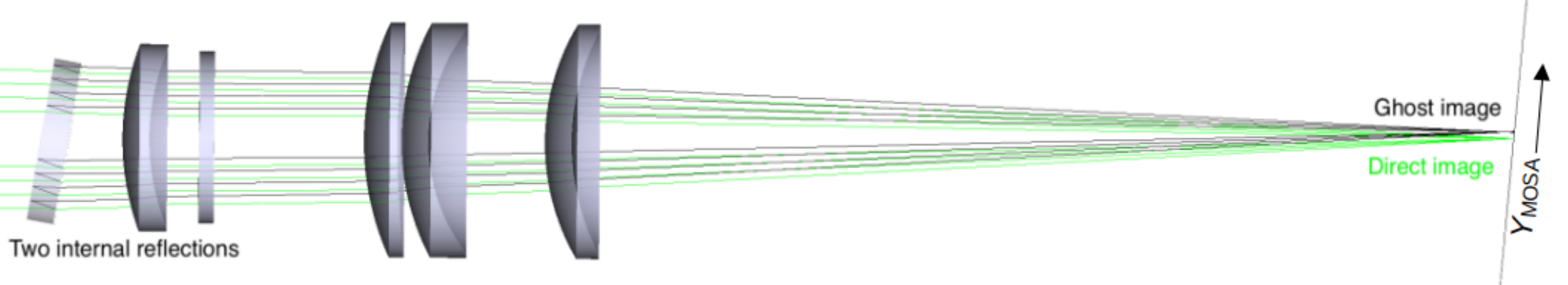}
\caption{{\tt Zemax} ray tracing. \textit{Top panel}: Shown is the nominal light path through the NISP optical elements; we note that the dichroic is actually a part of the telescope and not of NISP. The different colours show the light paths for three different sources in the field-of-view. For details of the optical layout and components see \citet{EuclidSkyNISP}. \textit{Bottom panel}: Selecting the middle (green) rays of the top panel, we show how the light is internally reflected inside the dichroic to produce a ghost that is offset from the source position. A similar double reflection happens in the filter to produce the filter ghost (suppressed in this plot for clarity).}
\label{fig:wave_path}
\end{figure*}

NISP delivers photometry in three bands (\YE, \JE, \HE) to an average $5\sigma$ point-source depth of about 24.4 \citep{EuclidSkyNISP} in the EWS. In \cref{fig:wave_path} we show the nominal light path through the NISP optical elements. Although the optics of Euclid have a complex interference coating layout for passband-forming and optical optimization, several parasitic reflections are still seen within the photometric (NISP-P) data. In this paper, we describe and model the two most common parasitic reflections in detail, that is the NISP dichroic and filter ghosts. 

Both features are caused by an internal double-reflection inside the respective optical element. The added optical path length -- of 2$\times$ the traversed element thickness in combination with the respective refractive index -- results in a defocused image of the ghosts on the Focal Plane Array (FPA). The considerable oblique angles of incidence (AOIs) in \Euclid's telescopic off-axis design \citep{Racca2016} is naturally responsible for the displacement of the dichroic ghosts from their source stars in the FPA (see \cref{fig:wave_path}; bottom). As these offsets are related to the incoming AOIs, we see variations across the FPA. We note here that VIS also has a dichroic ghost that is described in \cite{EuclidSkyVIS} and \cite{Q1-TP002}. However, the incoming AOIs are almost normal to the filter surface, so the displacement of the filter ghosts from their source stars arise mostly from the internal reflection off the  spherically convex surface of the
entrance side of the filter \citep{EuclidSkyNISP}.

Other unwanted effects not discussed in detail in this paper include persistence arcs from the filter wheel, arc-like reflections from bright stars likely caused by the NISP lenses, and glints from out-of-field bright stars. Brief descriptions and examples are shown in Figs.\ 19, 20, and 21 of \citet{EuclidSkyNISP}.

In \cref{sc:Data} we describe the data used to characterise the dichroic and filter ghosts; while the detection algorithms are described in \cref{sc:Dect}. We describe the characteristic of the dichroic and filter ghosts in \cref{sc:dichroic} and \cref{sc:filter} respectively. Finally, we conclude in \cref{sc:Con}. All magnitudes in this paper, unless stated otherwise, are in the AB system and all object magnitudes used for the ghost relations that are beyond the NISP saturation limit are transformed from \textit{Gaia} as in \citet{Q1-TP003}. On-sky sizes in figures are reported using a pixel scale of 0.3\arcsec pixel$^{-1}$ \citep{EuclidSkyNISP}.


\begin{figure}[t]
\centering
\includegraphics[width=1.0\hsize]{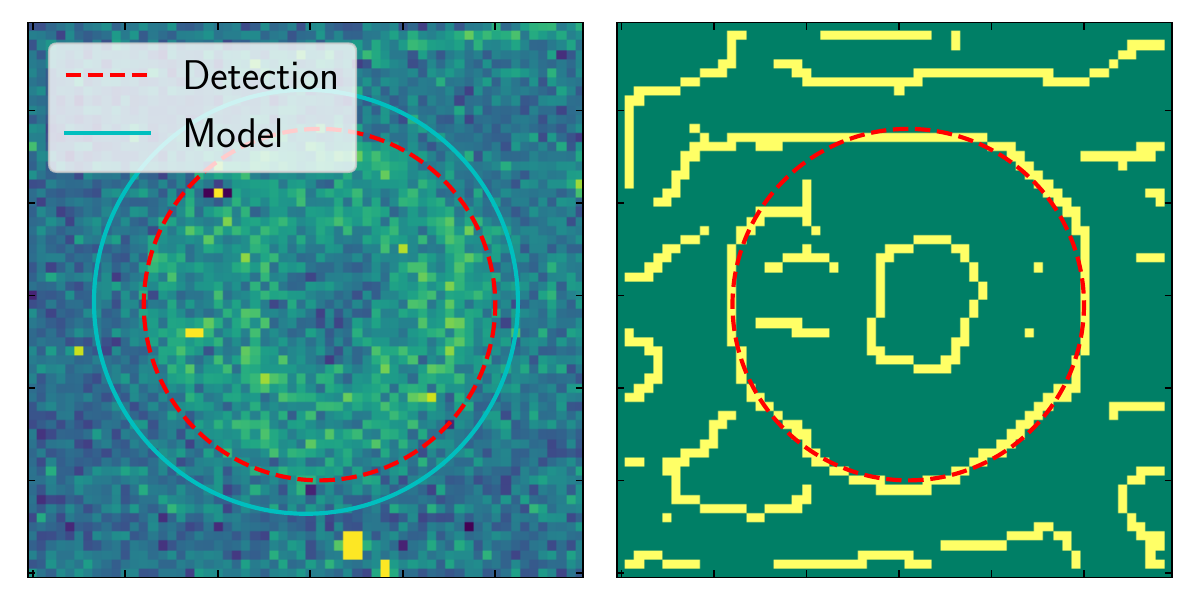}
\caption{Example of a dichroic ghost (\textit{left}, 60\,$\times$\,60 pixel -- 18\,$\times$\,18\arcsec\, -- cutout) and the automatic detection of the shape (red dashed line) based on edge detection using a circular Hough transformation (\textit{right}). Also shown is the final model (cyan solid line) used for Q1 based on the detection values and the radius derived based on the magnitude of the source star (see \cref{sc:dichroic}).}
\label{fig:di_detect}
\end{figure}


\section{Data}
\label{sc:Data}

To model the NISP dichroic and filter ghosts, we looked at processed NISP-P data from PV, with the most current version of the NIR pipeline (see \cite{Q1-TP003} for details) at the time of data extraction. In order to have bright stars in multiple positions on the detector to fully cover the FPA, we chose observations with a large number of widely dithered exposures. This includes (i) the self-calibration field \citep{EuclidSkyOverview} near the north ecliptic pole consisting of 60 dithers within a \ang{1;;} radius, (ii) observations of the \textit{Hubble} Space Telescope CALSPEC white dwarf GRW+70 5824 \citep{Bohlin2020} used for 
spectrophotometric calibration \citep{Q1-TP006}, placing the star at five positions on each detector, and (iii) the survey validation observations which consisted of visits of multiple different fields, including the Chandra Deep Field South \citep[CDFS;][]{Giacconi2001}, COSMOS \citep{Scoville2007}, the Euclid Deep Field North \citep[EDF-N;][]{EuclidSkyOverview}, GOODS-N \citep{Giavalisco2004}, and the Euclid Deep Field South \citep[EDF-S;][]{EuclidSkyOverview}.

This resulted in a total of 3270 images available for ghost detection, which contain 6023 instances where a suitably bright star falls onto the detectors.


\begin{figure*}[t]
\centering
\includegraphics[width=0.9\hsize]{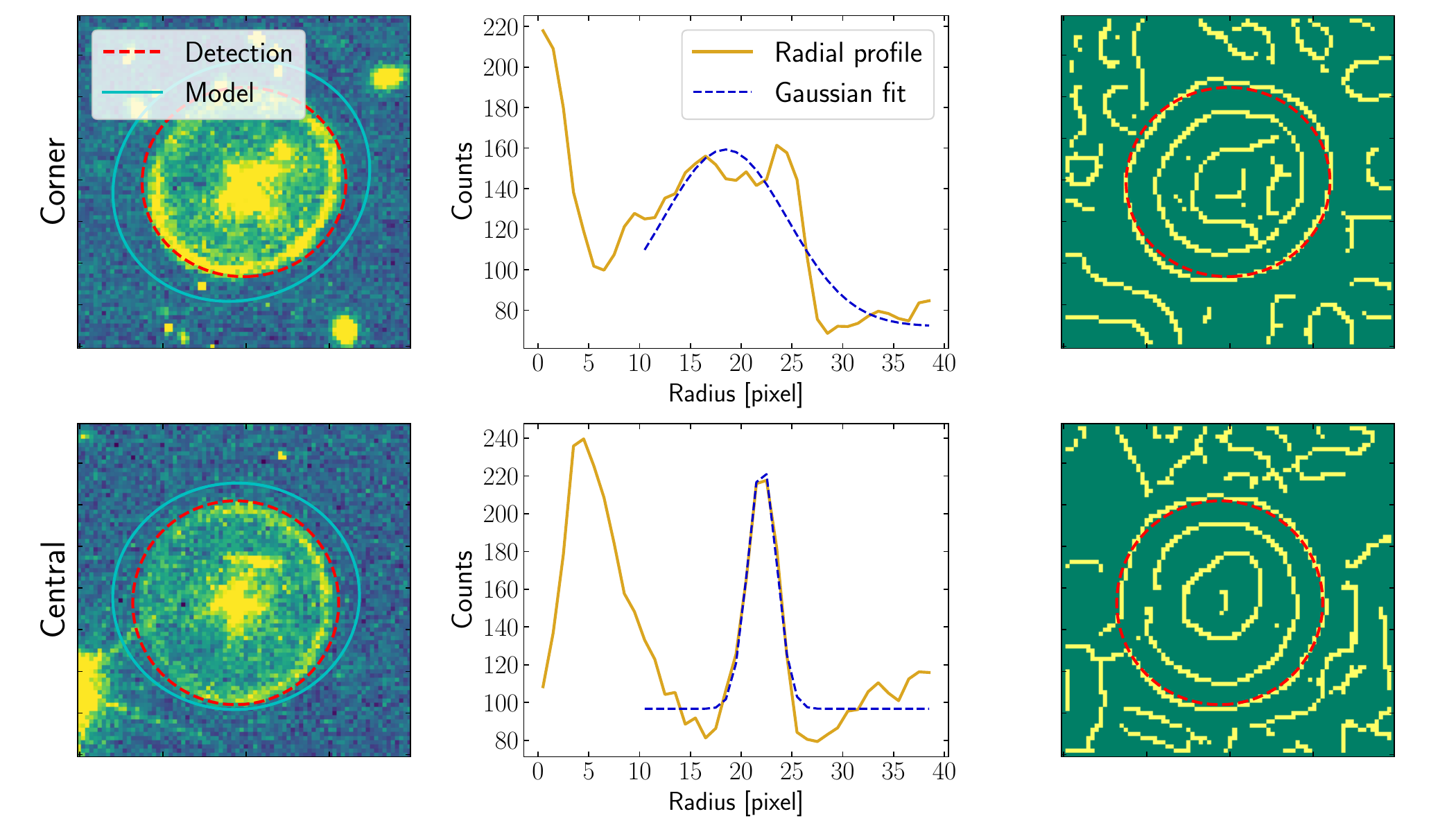}
\caption{Examples of filter ghosts near the FPA centre (\textit{bottom row}) and near a corner (\textit{top row}), highlighting the large field dependence. The \textit{left column} shows an 80\,$\times$\,80 pixel (24\,$\times$\,24\arcsec) cutout along with the automatic detection of the shape (red dashed line). Also shown is the final model (cyan solid line) used for Q1 based on the detection values, and the radius derived based on the magnitude of the source star (see \cref{sc:Dect}). The \textit{middle column} shows the total radial profile of the filter ghost cutout, and the Gaussian fit used as an initial guess for the detection radius. In the \textit{right column} we display the edge image created from the Canny filter and the automatic detection of the shape (red dashed line) based on an elliptical Hough transformation.}
\label{fig:fil_detect}
\end{figure*}

\section{Detection}
\label{sc:Dect}

During the NISP ground test campaign (which was performed on the NISP flight model and thus excludes the dichroic, see \cite{Gillard2025} for full details), a number of tests were undertaken inside a vacuum chamber to verify NISP’s optical performance. One such test provided a rough estimate of filter ghost images within NISP. All measurements were compared with Zemax simulations to validate and clarify the output data from the NISP system. This included a model of the dichroic ghost to estimate the expected shape and offsets of the dichroic ghosts. The main purpose of these dichroic ghost simulations were to enable the inclusion of realistic dichroic ghosts in the NISP simulations and to prepare the processing pipelines for ghost masking. As such, these simulations provided a good estimate for the dichroic ghost, but were not built to produce reliable models for direct comparison with real data.

Therefore, initial flight masking models for both the dichroic ghosts and filter ghosts were manually rebuilt using a limited number of self-calibration observations. Initially, for each observation each of the 16 H2RG detectors were treated separately, and the position of bright stars along with the positions of the resulting dichroic and filter ghosts within the detector were recorded. The offsets between the position of the bright star in FPA coordinates \cite[where the NISP pixel pitch is 18\,$\mu$m in both direction; ][]{EuclidSkyNISP} and the positions of the dichroic and filter ghosts was then used to build the initial 2D models to describe the ghost positions.

Next, a code to automatically find and determine the positions and other characteristics of the ghosts was used on all the available data. These characteristics were used to build a new ghost model. As better methods were developed and tested, this code was run multiple times, refining both the detection algorithm and the ghost model, which served as the initial guess for predicting ghost positions at each iteration. In this way, the model can be easily updated in the future once more data is available. As the model is built on FPA coordinates, it is independent of the detector positions and the gaps within the FPA footprint. As such, the code was also updated to work on bright star and ghost pairs that span across different detectors. From the data, it was found that only brighter stars with \YE, \JE, \HE $< 10$ produce dichroic and filter ghosts of concern, which must be masked out. Next, we describe the detection algorithm.

First, in order to find bright near-infrared stars in the footprint, for each observation, all stars with $J_{2MASS,\text{Vega}} < 8$ and within \ang{0.5} radius are queried from 2MASS \citep{Skrutskie2006}. For each star that falls onto a detector, the position of the star in pixel coordinates is determined using a 2D-Gaussian centroid and converted into FPA coordinates. Due to the saturated nature of these stars in the \Euclid data, the final magnitudes were obtained by transforming their \textit{Gaia} magnitudes to the \Euclid system \cite[to be consistent with the NIR pipeline, see ][]{Q1-TP003}.  Then, using the initial guess for the ghost position, 60\,$\times$\,60 pixel (80\,$\times$\,80 pixel) cutouts are created for the dichroic (filter) ghost. A 2D background for the cutouts is calculated by interpolating over a low-resolution background map. This low-resolution background map is created from the median within 10\,$\times$\,10 super pixels (resulting in 6\,$\times$\,6 and 8\,$\times$\,8 maps for the dichroic and filter ghost cutouts respectively); followed by applying a 3\,$\times$\,3 pixel sliding 2D median filter to suppress local under/over estimations. After background subtraction, astronomical sources with a minimum size of four pixels and with a signal exceeding a threshold based on the background root mean square (RMS) ($\times$2 for the dichroic and $\times$18 for the filter ghosts) are subtracted from the data by replacing the pixel values with the median background subtracted value.

To determine the position and radius of the ghosts, we first renormalise the cutouts using a 10\%, 90\% clipping to boost the contrast. We then perform edge detection using the Canny filter \citep{Canny1986} implemented in the {\tt skimage} Python package \citep{scikit-image}.
Afterwards, a circular Hough transformation \citep{Illingworth1987} is used to detect circular shapes in the edge image. The final radius of the ghosts and the corresponding central position are extracted from the Hough transformation, and converted to FPA coordinates for use in determining the models.

Due to the relatively constant radius of the dichroic ghost (see \cref{sc:dichroic}), the circular Hough transformation easily extracts the correct radius when using an initial estimate of 15--36 pixels (see \cref{fig:di_detect}). The filter ghosts are different, though. The thin ring present in them (i) is highly field-dependent in shape and size (see \cref{sc:filter}), and (ii) results in both an inner and outer radius in the Canny edge image. Therefore, the true radius of the filter ghosts are harder to automatically determine with a fixed radius range. To counter this, a Gaussian profile was fitted to the total radial profile of the filter ghost cutouts, excluding the central region containing the cusp (see \cref{fig:fil_detect}). The mean of the fitted Gaussian then provided an estimate of the radius for each filter ghost. These radii measurements were then used to build a simple model of the filter-ghost radius as a function of FPA position and then reincorporated into the code to allow a dynamic range for the Hough transformation. The initial radius range for the filter ghost is thus from $r_\mathrm{mod}-10$ to $r_\mathrm{mod} + 5$ pixels, where $r_\mathrm{mod}$ is the estimated radius from the simple model. 

Another complication arises from the non-circular (elliptical) shape of the filter ghost. Therefore, we use an ellipse Hough transformation \citep{Xie2002} to look for the ring. The results are then filtered by the minimum (20 pixels) and maximum (30 pixels for the minor and 35 pixels for major axis) values seen for the radii, as well as by a ratio of 1.3 between the minor and major axis, before finding the highest peak determined from the edge image. While the offsets of the filter ghost do not depend on the wavelength, the shape and brightness are wavelength dependent. Therefore, each filter needs to be treated separately in this analysis. Only the more compact filter ghosts in the \YE band data were easily detectable in an automatic way. Thus, the offsets of the filter ghost were only measured for the \YE band on the individual data, while the shape (radius and orientation) of each filter was found separately on the stacked data (see \cref{sc:filter}).


\begin{figure*}[t]
\centering
\includegraphics[width=0.48\hsize]{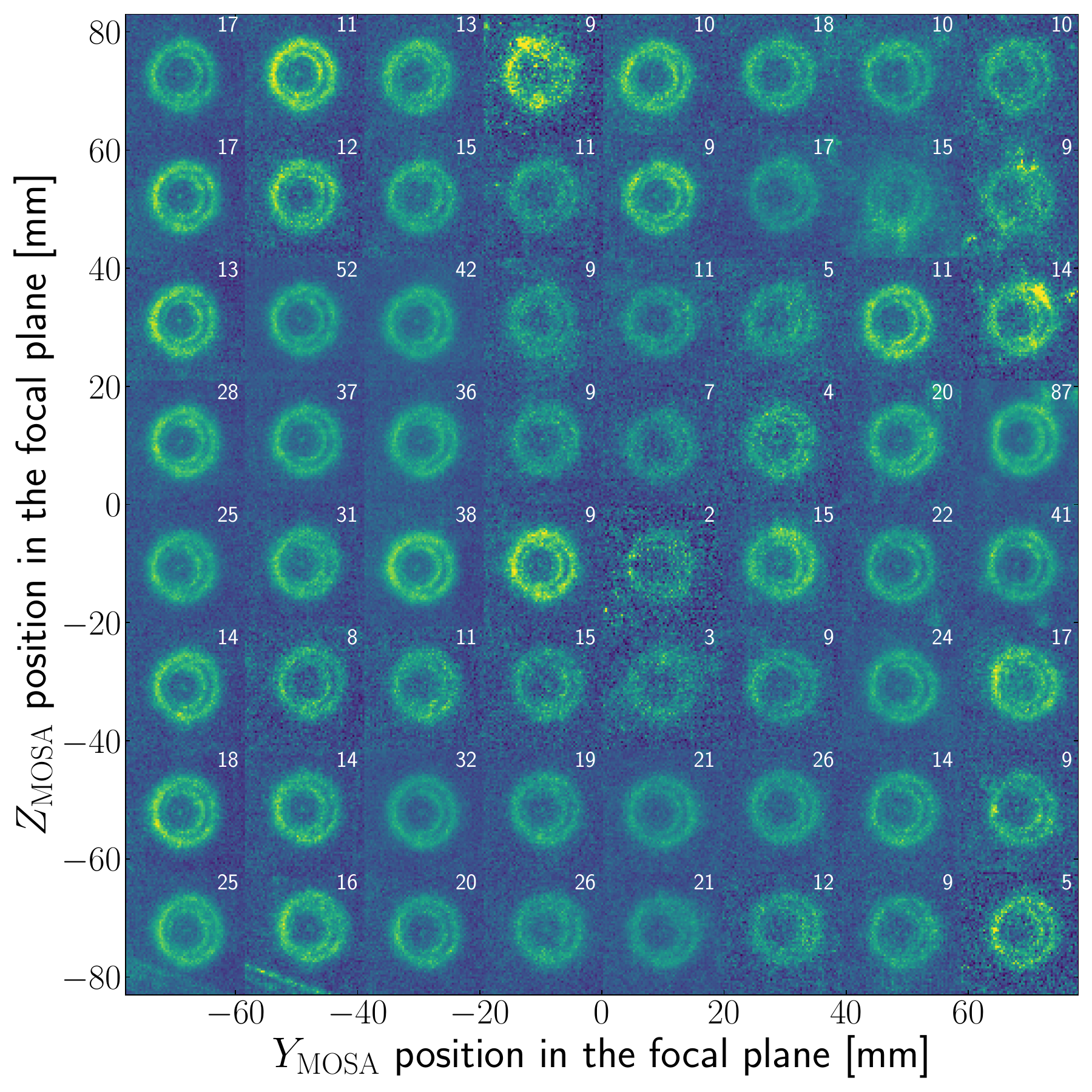}
\includegraphics[width=0.48\hsize]{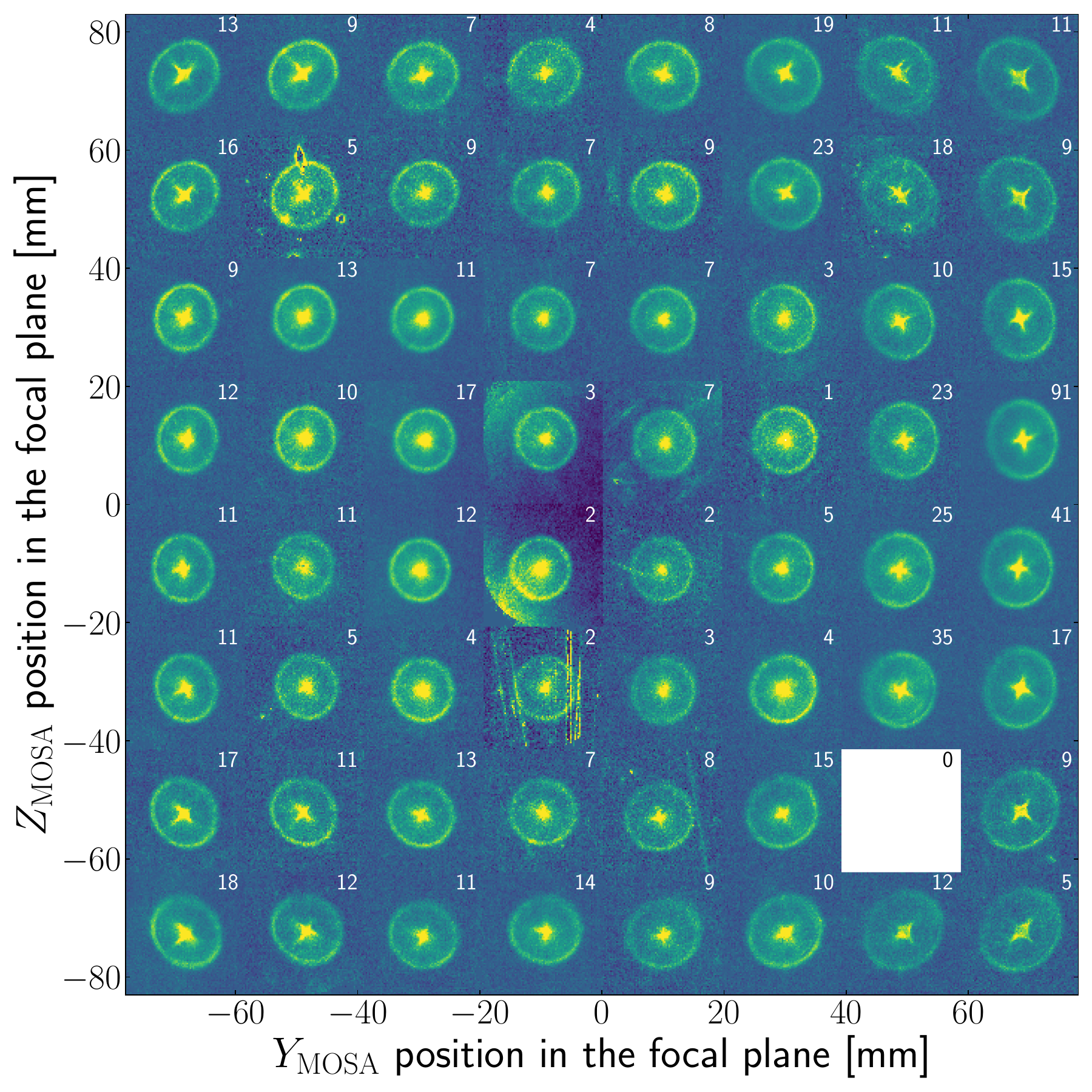}
\caption{Shape of the dichroic ghost (\textit{left panel}) and filter ghost (\textit{right panel}) as a function of field position. For the dichroic ghosts, images are made from a median combination of $60\,\times\,60$ pixel (18\,$\times$\,18\arcsec) cutouts based on a 20-fold binning on the FPA, restricted to a single magnitude bin in \HE band. For the filter ghosts, images are made from a median combination of 80\,$\times$\,80 pixel (24\,$\times$\,24\arcsec) cutouts on a 20-fold binning on the FPA, restricted to a single magnitude bin in the \YE band. The number of cutouts used to create each median is given in the top right of each median cutout. For the blank square we could not find suitable stars within that region of the FPA. Those showing artifacts such as arcs, streaks, or circular residuals resulting from inadequate masking in the individual images, are caused by having data from few ghosts.}
\label{fig:shape}
\end{figure*}

\begin{figure*}
\centering
\includegraphics[width=1.0\hsize]{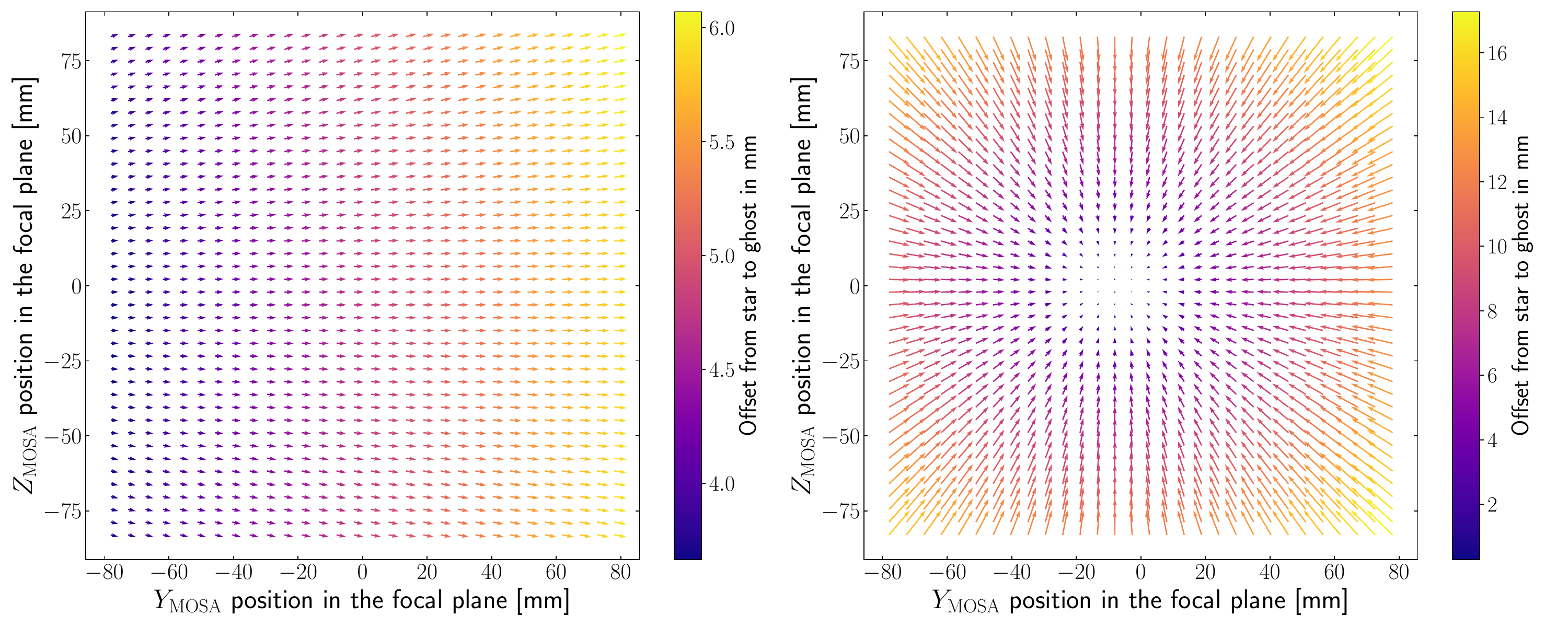}
\caption{Quiver plots showing the offset in millimetre from the source star to the dichroic (\textit{left}) and filter ghost (\textit{right}) across the FPA i.e. $-p_{\mathrm{Offset}}$. The matrices used to describe these models are given in \cref{ap:di} and \cref{ap:fil}.}
\label{fig:quiver}
\end{figure*}

\section{Dichroic ghost}
\label{sc:dichroic}

Here we describe the general characteristics of the NISP-P dichroic ghost. The shape of the dichroic ghost is characterised by a doubled-ringed doughnut with an off-centred hole due to \Euclid's off-axis design. The general shape and size  of the dichroic ghost changes very little as a function of FPA position (see \cref{fig:shape}; left).

\subsection{Offsets}
\label{sc:Offsets_di}
The offsets (defined as the difference between the centre of the circular fit determined for the dichroic ghost and the central position of the source star; in mm) in the $Y_\mathrm{MOSA}$ and $Z_\mathrm{MOSA}$ axes for the dichroic ghost are both well described by 3rd-order polynomials in the form of
\begin{equation}\label{eq:poly}
    p_{\mathrm{Offset}}(Y_\mathrm{MOSA},Z_\mathrm{MOSA}) = \sum_{i,j} c_{i,j} * Y_\mathrm{MOSA}^i * Z_\mathrm{MOSA}^j,
\end{equation}
where $p_{\mathrm{Offset}}$ provides the offset between the central positions of the dichroic ghost and the bright star pair in a given axis and $c_{i,j}$ is the matrix describing the coefficients of the polynomial for the offset in that axis. These matrices for the model used for \Euclid's first Quick Data Release (Q1) are given in \cref{ap:di}. As described in \cref{sc:Intro}, the displacement of the dichroic ghost on the FPA arises from the oblique AOIs from \Euclid's off-axis design. The total offset between the source star and the dichroic ghost ranges from 3--6\,mm (167--335 pixels; see \cref{fig:quiver}; left). The larger component is in the $Y_\mathrm{MOSA}$ axis is always positive when moving from the source star to the dichroic ghost position, while the offset in the $Z_\mathrm{MOSA}$ axis is smaller but ranges from negative to positive. The RMS for the Q1 model is 0.082\,mm (4.6 pixels).

\begin{figure}[t]
\centering
\includegraphics[width=1.0\hsize]{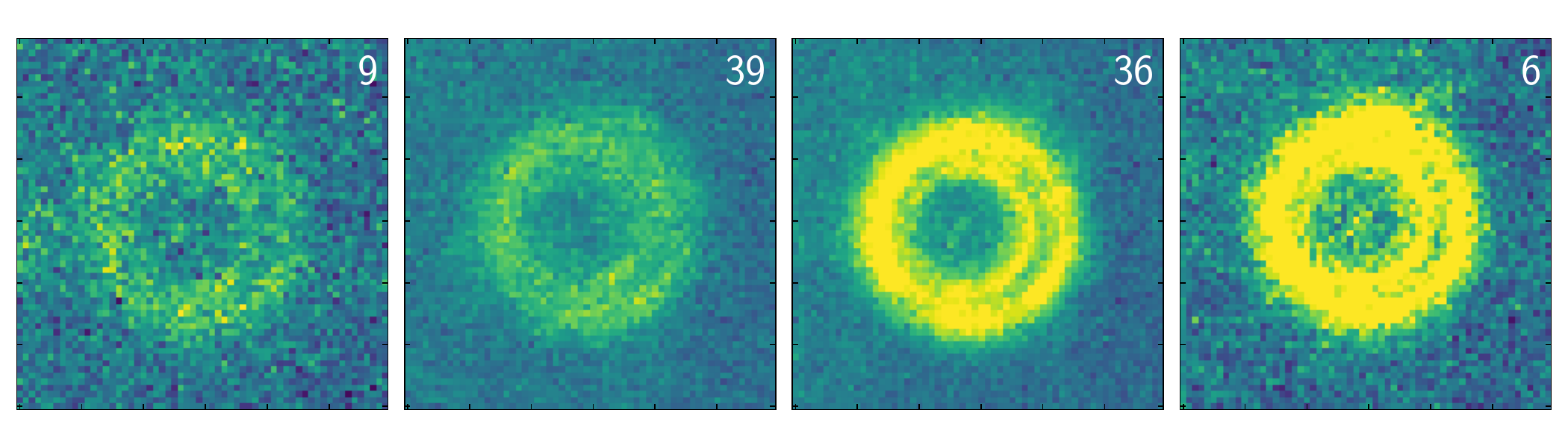}
\includegraphics[width=1.0\hsize]{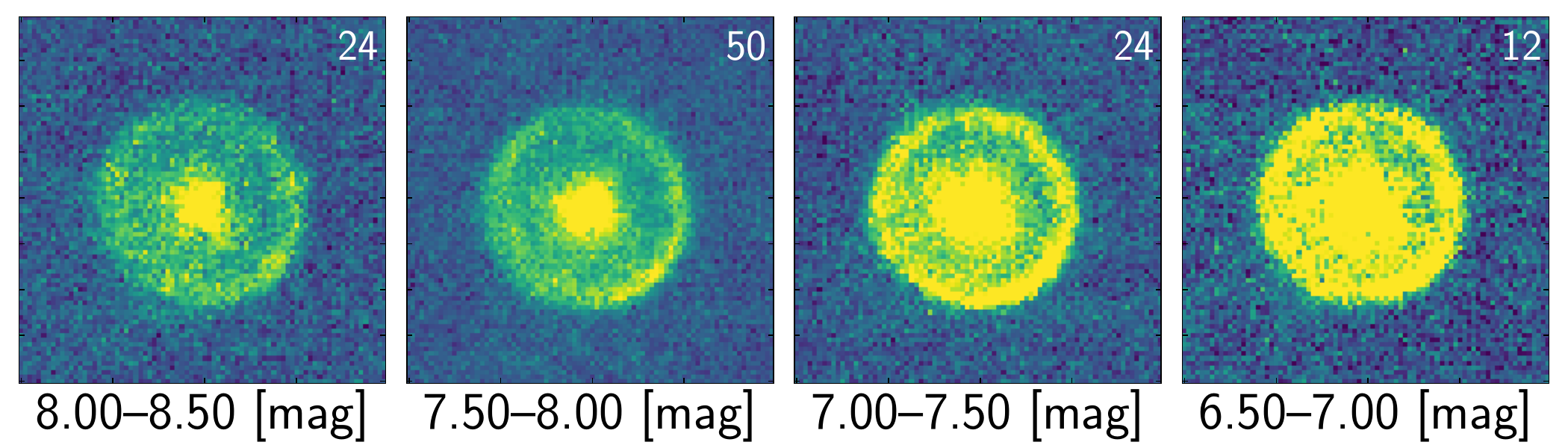}
\caption{Appearance of the dichroic ghost (\textit{top row}) and filter ghost (\textit{bottom row}) as a function of source-star brightness. For the dichroic ghosts, images are made from a median combination of 60\,$\times$\,60 pixel (18\,$\times$\,18\arcsec) cutouts, restricted to one region of the FPA for the \HE band. For the filter ghosts, images are made from a median combination of 80\,$\times$\,80 pixel (24\,$\times$\,24\arcsec) cutouts, restricted to the central region of the FPA for the \YE band. The number of cutouts used to create each median is given in the top right of each median cutout.}
\label{fig:flux}
\end{figure}

\subsection{Magnitude relations}
\label{sc:Rad_di}

Due to slight differences in the internal structure, the dichroic ghosts cannot be modelled and subtracted, and thus must be masked to avoid contamination of photometric measurements. To meet the top level requirement of a relative photometric error below 1.5\% \citep{EuclidSkyOverview}, the detection-chain error must be smaller than 1\% \citep{CalCD-B}. Using the error estimates from \cite{CalCD-B} and the targeted 5$\sigma$ point-source depth in the EWS, it was calculated that this requirement  equates to a maximum contribution from a ghost of 0.0866\,e$^-$\,pixel$^{-1}$.

\begin{figure}[t]
\centering
\includegraphics[width=1.0\hsize]{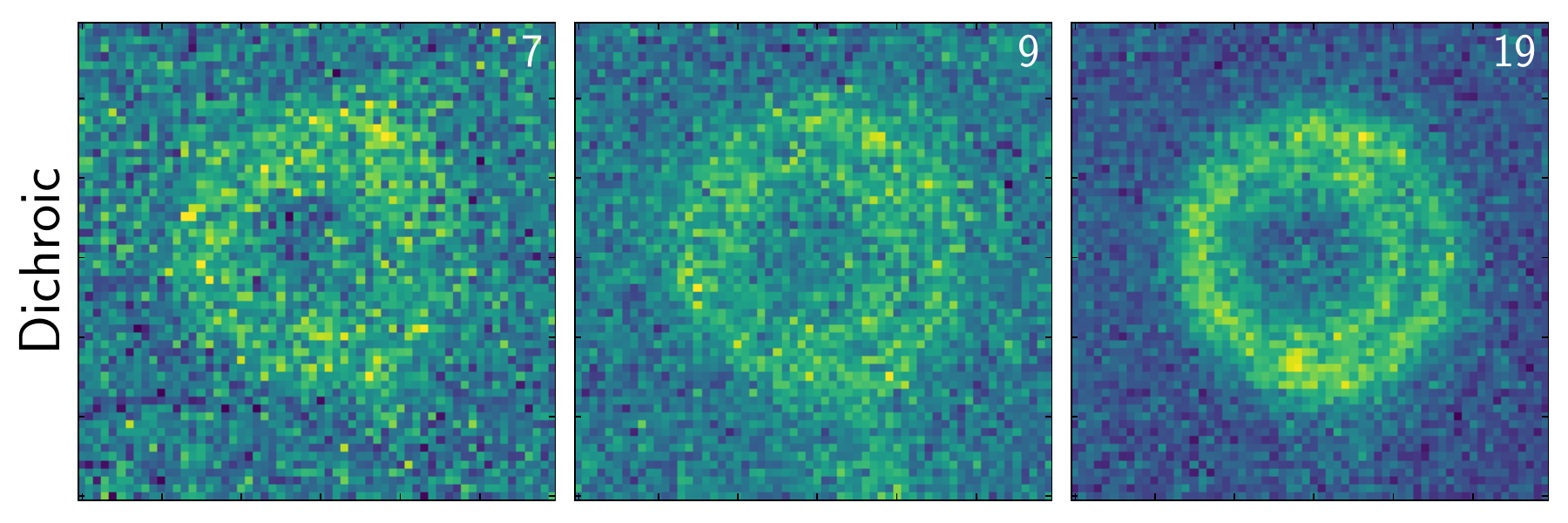}
\includegraphics[width=1.0\hsize]{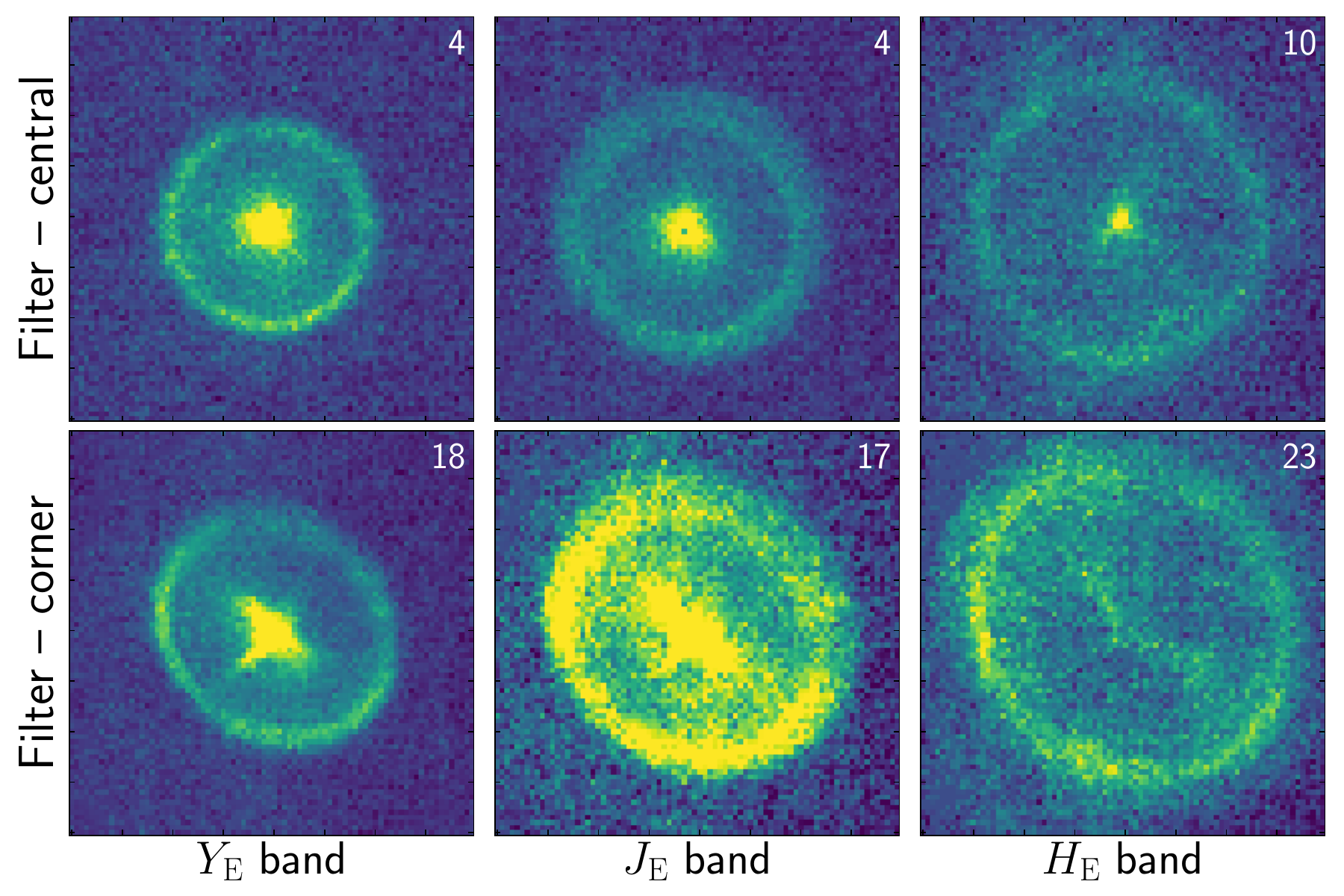}
\caption{Shape of the dichroic ghost (\textit{top row}) and filter ghost (\textit{lower rows}) as a function of waveband. For the dichroic ghosts, images are made from a median combination of 60\,$\times$\,60 pixel (18\,$\times$\,18\arcsec) cutouts, restricted to one region of the FPA and a single magnitude bin. For the filter ghosts, since there is a dependency of the shape on the FPA position, images are made from a median combination of 80\,$\times$\,80 pixel (24\,$\times$\,24\arcsec) cutouts, restricted to either the central region (middle) or a corner region (bottom) of the FPA and a single magnitude bin. The number of cutouts used to create each median is given in the top right of each median cutout.} 
\label{fig:wave}
\end{figure}

\begin{figure*}[t]
\centering
\includegraphics[width=0.48\hsize]{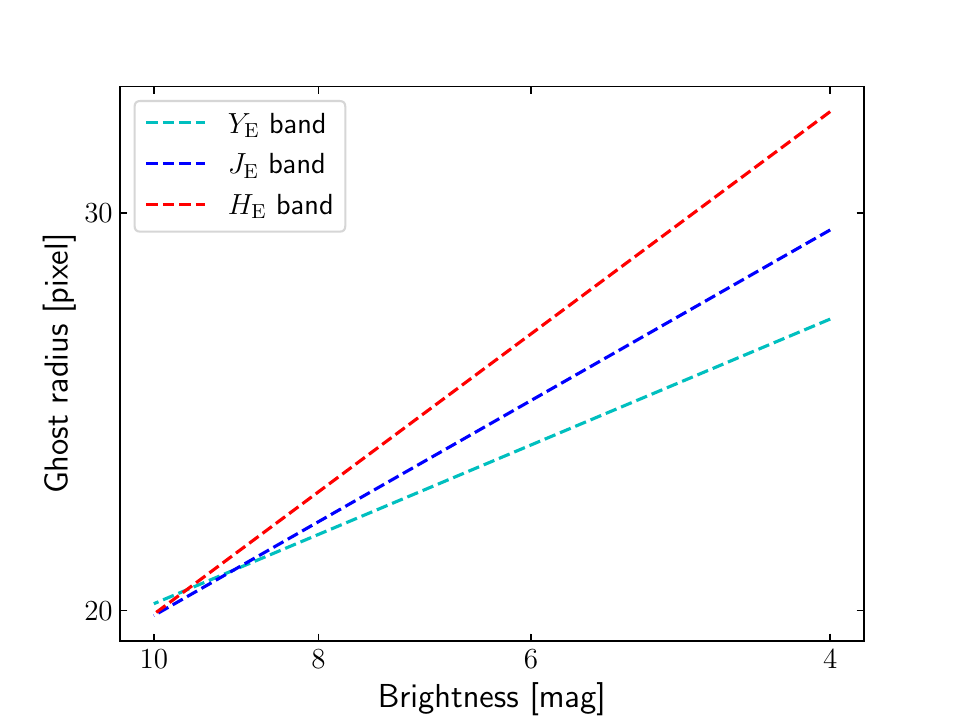}
\includegraphics[width=0.48\hsize]{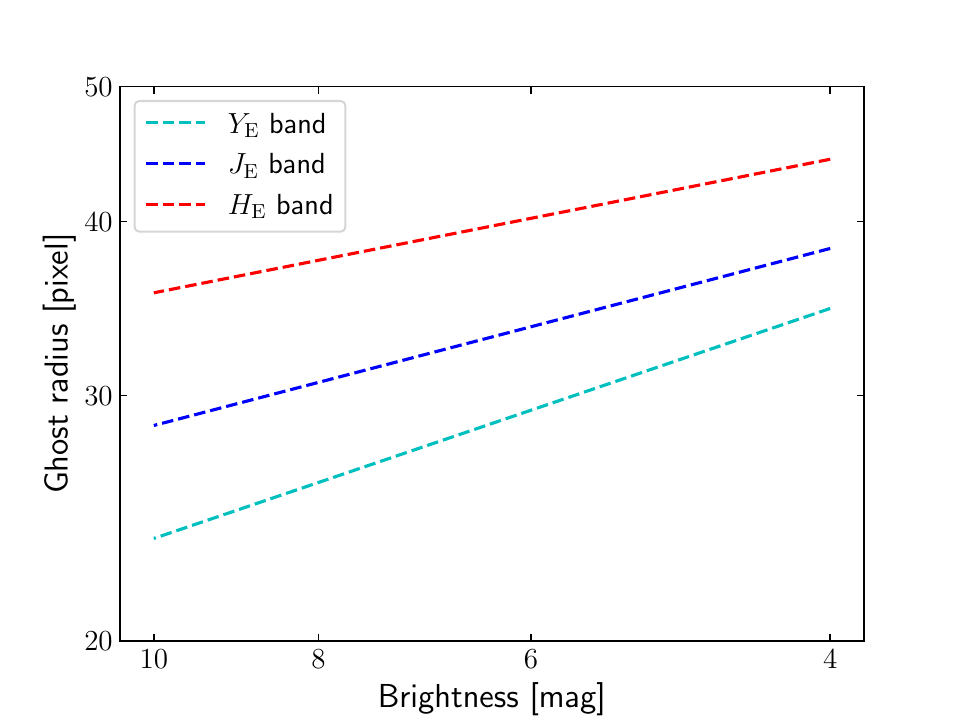}
\caption{The radius-magnitude relation, given by the coefficients in \cref{table:rad}, for each filter for the dichroic ghost (\textit{left}) and filter ghost (\textit{right}); described in \cref{sc:Rad_di}.}
\label{fig:rad_mag}
\end{figure*}

To determine the radius within which we need to mask the dichroic ghosts, $R_{\rm mask}$, we need to determine at which radius its flux falls below the maximum contribution allowed from ghosts set by the requirement. Since the surface brightness of the dichroic ghost is dependent on the magnitude of the source star (see \cref{fig:flux}; top), as well as the wavelength it is observed in (see \cref{fig:wave}; top), each filter is treated separately while combining ghosts from source stars with similar magnitudes. For each dichroic ghost, we constructed 100 radial profiles starting from the centre position of the dichroic ghost and spread over the entire circumference to a radius of 50 pixels. A sigma-clipped median of these radial profiles was then calculated for multiple dichroic ghosts with similar source-star magnitudes to create a smooth radial profile. After subtracting any residual background -- calculated from the median of the radial profile at a radius greater than 30 pixels -- the radius at which the flux falls below the requirement is recorded. All measurements are then used to fit a power law, in the form of 
\begin{equation}\label{eq:radius}
    R_{\rm mask} = R_0 m^{\Gamma},
\end{equation}
where $R_0$ describes a radius in pixels and $m$ the source star magnitude. The coefficients found and used for Q1 are given in \cref{table:rad}. The relation for each filter is shown in \cref{fig:rad_mag} (left). These relations are used to compute the radius of the dichroic-ghost mask based on the magnitude of the source star.

\begin{figure*}[t]
\centering
\includegraphics[width=0.48\hsize]{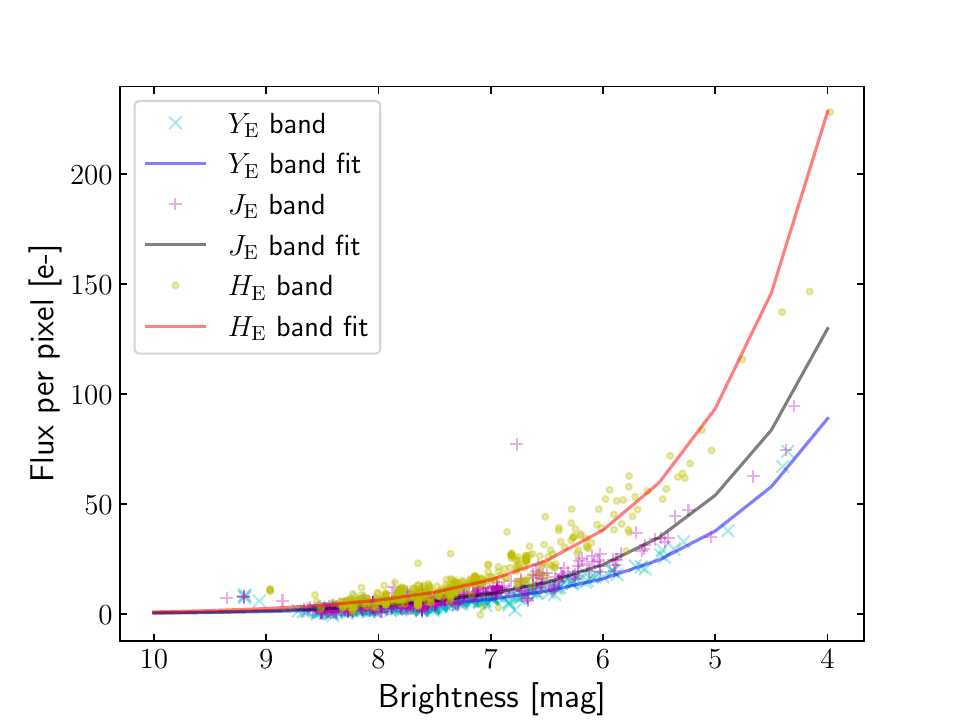}
\includegraphics[width=0.48\hsize]{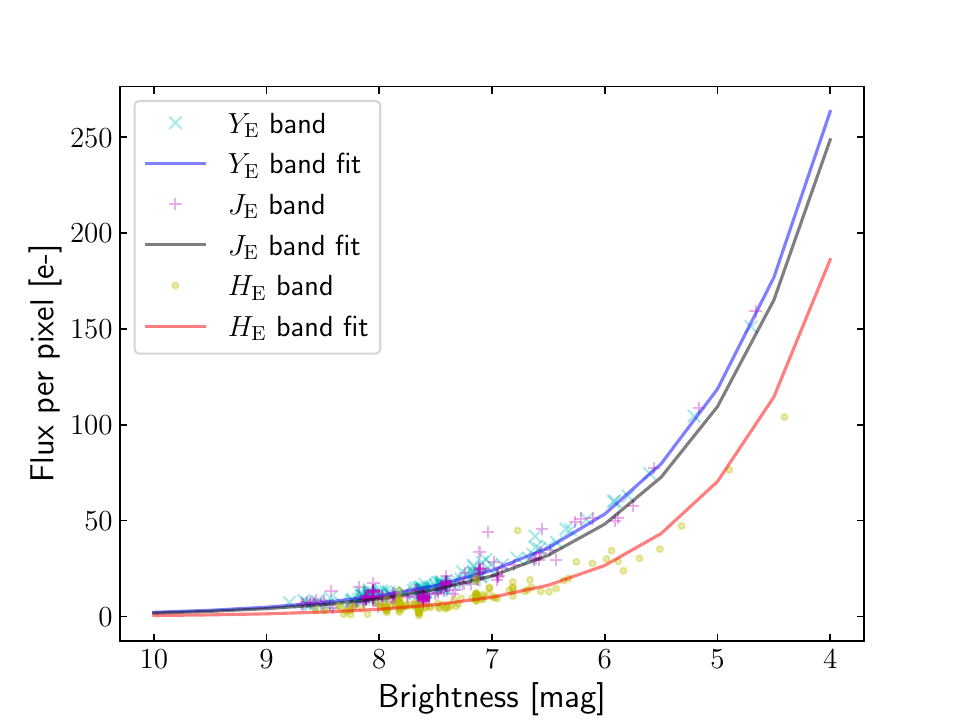}
\caption{The mean flux per pixel in the dichroic (\textit{left}) and filter (\textit{right}) ghost, calculated from the smoothed radial profiles as described in \cref{sc:Rad_di}. The minimum brightness for masking is thus determined for each filter by the requirement that the ghost must contribute less than 0.0866\,e$^-$\,pixel$^{-1}$.}
\label{fig:mag}
\end{figure*}

To determine the minimum brightness of a source star that would create a dichroic ghost that requires masking ($m_\mathrm{min}$), we compute the surface brightness of the dichroic-ghost profiles calculated in the previous step. All flux within the previously calculated radius is included. A straight line is then fit to the mean flux per pixel in the dichroic ghost versus the flux from the source star. The magnitude of the source star at which the dichroic-ghost flux reaches the requirement is then assigned as the $m_\mathrm{min}$ for the respective filter. The values used for Q1 are given in \cref{table:rad}. These relations, as a function of source-star magnitude, are shown in \cref{fig:mag} (left).

The NISP ground tests initially estimated the ratio of the peak flux in the source star to the peak surface brightness of filter ghosts to be $8\times 10^{-8}$. Using the NISP-P model PSF and zero points from \cite{EuclidSkyNISP}, we estimate the peak flux from the magnitude of the source star and find dichroic ghost ratios of $5.8\times 10^{-9}$, $5.7\times 10^{-9}$, and $8.3\times 10^{-9}$ for \YE, \JE, and \HE\ respectively. These are an order of magnitude below the initial estimates, exceeding the predictions and highlighting the optical performance of NISP. 

\begin{table*}[htbp!]
\centering
\caption{Q1 ghost masking parameters.}
\smallskip
\label{table:rad}
\smallskip
\begin{tabular}{c|D{,}{\pm}{-1}D{,}{\pm}{-1}D{,}{\pm}{-1}|D{,}{\pm}{-1}D{,}{\pm}{-1}D{,}{\pm}{-1}}
\hline
\noalign{\vskip 1pt}
  & \multicolumn{3}{c|}{Dichroic ghost} & \multicolumn{3}{c}{Filter ghost} \\
\hline
\noalign{\vskip 1pt}
Filter & \multicolumn{1}{c}{\hspace{2pt}$R_0$} & \multicolumn{1}{c}{\hspace{2pt}$\Gamma$} & \multicolumn{1}{c|}{\hspace{2pt}$m_\mathrm{min}$} & \multicolumn{1}{c}{\hspace{2pt}$R_0$} & \multicolumn{1}{c}{\hspace{2pt}$\Gamma$} & \multicolumn{1}{c}{\hspace{2pt}$m_\mathrm{min}$}\\[1pt]
  \hline
\noalign{\vskip 1pt}
\YE & 41.77,2.36 & -0.317,0.028 & 9.02,0.33 & 61.56,8.99 & -0.415,0.07 & 10.74,0.27\\
\JE & 53.46,3.06 & -0.429,0.028 & 9.27,0.31 & 60.05,6.01 & -0.324,0.05 & 10.49,0.46\\
\HE & 72.25,4.26 & -0.560,0.030 & 9.83,0.27 & 61.91,8.77 & -0.241,0.07 & 9.17,0.64\\
\hline
\end{tabular}
\tablefoot{Coefficient results ($R_0$ and $\Gamma$) for the radius-magnitude relation (see \cref{sc:Rad_di} and \cref{fig:rad_mag}) and the minimum masking brightness, in magnitudes, ($m_\mathrm{min}$) for the dichroic and filter ghosts used for Q1.}
\end{table*}

\begin{figure*}[t]
\centering
\hspace*{-1.9cm}
\includegraphics[width=1.2\hsize]{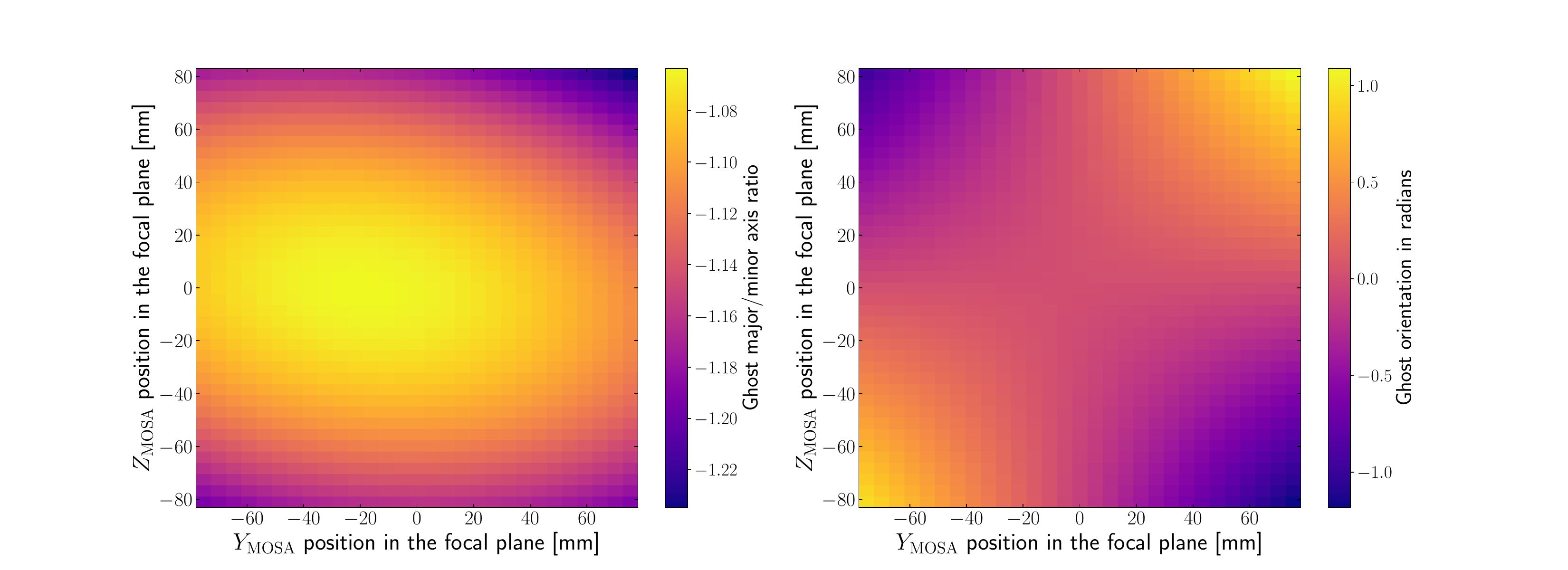}
\caption{Filter ghost shape and orientation. \textit{Left:} Model of the filter ghost's major-to-minor axis ratio, that is the elongation. \textit{Right}: Model of the orientation of the filter ghost in radians, measured clockwise from the $Y_\mathrm{MOSA}$-axis. The matrices used to describe these models are given in \cref{ap:fil}.}
\label{fig:fil_rat_or}
\end{figure*}

\section{Filter ghost}
\label{sc:filter}

Here we describe the general characteristics of the NISP-P filter ghost. The shape of the filter ghost varies much more depending on the position on the FPA (see \cref{fig:shape}; right). The variations show a radially symmetric pattern, with the centre of the variations off-set from the centre of the FPA at roughly $Y_\mathrm{MOSA} = -10$\,mm. The general shape consists of a thin ring and a central cusp. For small offsets from the pattern centre/off-centred FPA position, the ring is circular in shape and the cusp is point-like, while larger radial offsets result in elongation of the ring and the central cusp. Along the central $Y_\mathrm{MOSA}$ and $Z_\mathrm{MOSA}$ axis, the elongation of the ring is small while the central cusp changes into a 4-point star-like shape with a radial axis and axis perpendicular to the radial component. Towards the corners, the elongation is more pronounced, resulting in an oval shape; while the central cusp elongates much more in the plane perpendicular to the radial axis, becoming one-sided and resulting in a `bird-like' feature. These changes are the result of the geometric distortion introduced by the spherically convex surface of the entrance side of the filters. The curvature results in slightly different path lengths for the light during the second internal reflection, thus affecting the final image of the filter ghost. The size of the filter ghost, as well as the sharpness of the ring and central cusp, is also dependent on wavelength, with the redder filters producing a more out-of-focus image (see \cref{fig:wave}; middle and bottom). This is due to the different filter thicknesses, which in turn result in slightly different path lengths and thus focal points on the detector. Internal structures within the filter ghosts (multiple thin concentric rings) are seen only for the brightest stars up to \YE, \JE, \HE\ $>$ 4 \cite[brighter stars are avoided by the survey, ][]{EuclidSkyOverview}.

Spatial variation of the ratio between the major and minor axis of the filter ghost -- describing the elongation -- is modelled by a 2nd-order polynomial, $p_{\mathrm{Radius}}$ in the same form as \cref{eq:poly}.  The model shows the off-centred, radially dependent pattern, with the filter ghost appearing circular towards the centre of the pattern and more elongated, up to a ratio of 1.22, towards the corners (see \cref{fig:fil_rat_or}; left). The orientation (in radians) of the elongation is also modelled by a 2nd-order polynomial, $p_{\mathrm{Orientation}}$, of the same form. There is no rotational preference (set to 0) along the central axes, where the shape is more circular (see \cref{fig:fil_rat_or}; right). Diagonal symmetry -- that is between opposite corners -- is seen with increasing negative rotation to the bottom left and increasing positive rotation towards the bottom right. The rotation is measured in radians from the $Y_\mathrm{MOSA}$-axis in a clockwise direction, with a maximum of just over 1\,radian (about 60\degree) seen in both directions. The matrices containing the coefficients for these two models used for Q1 are given in \cref{ap:fil}. The RMS for the ratio and orientation Q1 models is 0.23 and 0.63\,radians respectively.

\subsection{Offsets}
\label{sc:Offsets_fil}
The offsets (defined as the difference between the centre of the elliptical fit determined for the filter ghost and the central position of the source star; in mm) in the $Y_\mathrm{MOSA}$ and $Z_\mathrm{MOSA}$ axes for the filter ghost are also both well described by 3rd-order polynomials in the same form as \cref{eq:poly}. The matrices for the model used for Q1 are given in \cref{ap:fil}. Similar to the shape variations, the offset for the filter ghost from the source star greatly varies across the FPA, ranging between 1\,mm (56 pixels) and 17\,mm (949 pixels), with the same off-centred radially symmetric pattern (see \cref{fig:quiver}; right). Near the pattern centre/off-centred FPA position, the offset is as little as a few millimetres, while at greater radial positions the offsets are as large as 17\,mm. The offset to the filter ghost from the source star is always positioned towards the pattern centre/off-centred FPA position. Again, these variations are due to the spherically convex surface of the entrance side of the filter. The curvature results in slightly different AOIs for the light during the second internal reflection, while the concave shape (as see from within the filter) results in the light path bending inwards, changing the offset of the filter ghost in a radially dependent way. The RMS for the Q1 model is 0.045\,mm (2.5 pixels).



\subsection{Magnitude relations}
\label{sc:Rad_fil}

The surface brightness of the filter ghost is also dependent on the magnitude of the source star (see \cref{fig:flux}; bottom), as well as the wavelength it is observed in (see \cref{fig:wave}; middle and bottom). Thus, to determine $R_{\rm mask}$ for the filter ghosts, we followed the same approach as done for the dichroic ghost described in \cref{sc:Rad_di}, however, given that the radius and elongation of the filter ghosts not only depends on the wavelength, but also on the position on the FPA, only filter ghosts from the central region, where the shape is circular, are used to determine the radius-magnitude relation. The coefficients found and used for Q1 for \cref{eq:radius} are given in \cref{table:rad}. The relation for each filter is shown in \cref{fig:rad_mag} (right). These relations are used to compute the minor axis of the filter-ghost mask based on the magnitude of the source star. The major axis of the filter-ghost mask, which depends on the position on the FPA, is then obtained using the ratio model described in \cref{sc:filter}. The details on how this is done are described in \cref{ap:fil}.

To determine $m_\mathrm{min}$ for the filter ghost we follow the same approach as done for the dichroic ghost described in \cref{sc:Rad_di}, again restricting the calculation to the central regions where the filter ghost is circular in shape. The values used for Q1 are given in \cref{table:rad}, with the data and fits shown in \cref{fig:mag} (right).

Calculating the same source-star peak flux to ghost brightness ratios for the filter ghosts, as done for the dichroic ghosts in \cref{sc:Rad_di}, we find similar values for the filter ghost, with ratios of $1.3\times 10^{-8}$, $8.5\times 10^{-9}$, and $4.6\times 10^{-9}$ for \YE, \JE, and \HE\ respectively.


\section{Discussion}
\label{sc:Con}

As the data used to calculate these ghost models were from PV, they occurred before the first partial decontamination in March 2024. This decontamination was initiated after noticeable throughput change and resulted in two rounds of partial decontamination by warming up specific mirror surfaces during the subsequent period of heavy ice contamination between April and June 2024. An analysis of the dichroic ghost using data just before and after the March 2024 decontamination was performed as a sanity check, to ensure no significant shifts in the ghost position occurred due to the decontamination, possibly due to changes in the alignment of the optics. No significant shifts between the data before and after the decontamination were found for the dichroic ghost. An analysis of the throughput of the dichroic and filter ghosts within NISP-P due to ice contamination was not conducted within the scope of this project, and will be discussed elsewhere.

In this paper, we describe the characteristics and detection of the dichroic and filter ghosts in NISP-P images. For the detection, we developed an algorithm to automatically find and measure the positions of the dichroic and filter ghosts on NISP-P data. We present the models used to describe the offsets, the radius, the shape, and the brightness of the dichroic and filters ghosts. The values found from selected PV data for the coefficients used in the models are given in \cref{table:rad}, \cref{ap:di}, and \cref{ap:fil}.

From these data, we find a small range of offsets (3--6\,mm or 167--335 pixels) for the dichroic ghosts, with a much larger range of 1--17\,mm (56--949 pixels) for the filter ghosts. The shape of the dichroic ghosts is mostly circular and changes very little across the FPA or between filters. The radius of the circular dichroic-ghost mask ranges from 21 to 34 pixels based purely on the magnitude of the source star. The shape of the filter ghosts however, depends greatly on both their position on the FPA and the filter of the observation. The shape is elliptical in nature and thus requires an elliptical mask, with the minor axis ranging from 23 to 45 pixels based on the magnitude of the source star, and the major axis up to 1.22 times greater than the minor axis depending on the position on the FPA. The surface brightness of both the dichroic and filter ghosts are roughly an order of magnitude below the initial exceptions from the NISP ground tests, with only stars with magnitudes $<$ 9--11 producing ghosts needing to be masked; highlighting the on-sky performance of NISP. On average, the dichroic ghosts are fainter than the filter ghosts in the same filter, but different relations between the surface brightness with respects to the wavelength are observed. 

These models are used to mask affected pixels, where only ghosts resulting from stars and not extended sources which may be brighter than the magnitude limit are considered, in the Q1 data release. While a description of how to use the matrices presented in this paper correctly, with examples using Python, is provided in \cref{ap:di} and \cref{ap:fil}; the details of the ghost-masking routine within the NIR pipeline is described in \cite{Q1-TP003}. The further use of these masks during the stacking and catalogue creation is described in \cite{Q1-TP004}. For the Q1 data release, we find the total area lost due to masking of ghosts, persistence arcs, saturated stars, cosmic rays, dead pixels and detector gaps is $\sim$4\%. This percentage ranges between 2--10\% over the Q1 footprint with varying stellar density. Even for the highest stellar density regions, this number falls below the error budget of 12\% for area lost due to these effects in order for Euclid to comply with the science goals of the mission. As the survey continues, and more data become available, the models will be recomputed to update these parameters. However, we do not expect significant changes in the models over the lifetime of the mission resulting from optic shifts or ageing damage due to changes in the optics' coatings. While the persistence arcs from the filter wheel are masked in Q1 data \citep{Q1-TP003}, further work to mask the other unwanted optical effects seen within the NISP-P data, such as the arc-like reflections from the NISP lenses and glints from out-of-field bright stars, is only planned for future data releases.


\begin{acknowledgements}
The authors at MPIA acknowledge funding by the German Space Agency DLR under grant number 50~QE~2303.\\
We thank Pierre-Antoine Frugier for providing useful comments to improve the publication.\\
The plots in this publication were prepared with {\tt Matplotlib} \citep{hunter2007}.

\AckEC
\end{acknowledgements}


\bibliography{Euclid,Q1,ghosts}


\begin{appendix}
\onecolumn
\section{Dichroic ghost}\label{ap:di}
The 3rd-order polynomials used to describe the dichroic ghost offsets (from the dichroic ghost position to the source star position) are given by \cref{eq:poly}. The matrices containing the coefficients for the dichroic ghost offset in each axis, used for the Q1 models, are

\begin{align}
\mathrm{Offset}_{Y_\mathrm{MOSA}} &= 
\begin{pmatrix*}[l]
-4.75805908 & -4.08494811\times10^{-4} & -1.68202062\times10^{-6} \\
-1.41684848\times10^{-2} & +1.26835014\times10^{-7} & -1.07871607\times10^{-8}\\
-2.42159141\times10^{-6} & +2.76254122\times10^{-8} & +5.34023160\times10^{-11} \\
\end{pmatrix*}~\mathrm{and}\\[5mm]
\mathrm{Offset}_{Z_\mathrm{MOSA}} &= 
\begin{pmatrix*}[l]
-1.67583495\times10^{-1} & -1.42131358\times10^{-2} & +1.63649376\times10^{-7}\\
-3.01789225\times10^{-5} & -3.30794149\times10^{-6} & +1.93109689\times10^{-9}\\
+1.67513242\times10^{-7} & -5.36299455\times10^{-9} & -5.05874463\times10^{-11}\\
\end{pmatrix*}.
\end{align}

From these, the position of the dichroic ghost on the FPA (in mm) is given by:
\begin{equation}\label{eq:offsets}
    (Y_\mathrm{MOSA},Z_\mathrm{MOSA})_{\mathrm{dichroic}} = (Y_\mathrm{MOSA},Z_\mathrm{MOSA})_{\mathrm{star}} - (p_{\mathrm{Offset}_{Y_\mathrm{MOSA}}}(Y_\mathrm{MOSA},Z_\mathrm{MOSA}),p_{\mathrm{Offset}_{Z_\mathrm{MOSA}}}(Y_\mathrm{MOSA},Z_\mathrm{MOSA}))
\end{equation}.

The offsets themselves can be easily calculated in Python using the polyval2d function within numpy.polynomial.polynomial e.g.:
\begin{equation}\label{eq:poly_python}
    p_{\mathrm{Offset}_{Y_{\mathrm{MOSA}}}}(Y_\mathrm{MOSA},Z_\mathrm{MOSA}) = \mathrm{numpy.polynomial.polynomial.polyval2d}(Y_\mathrm{MOSA},Z_\mathrm{MOSA},\mathrm{Offset}_{Y_\mathrm{MOSA}})
\end{equation}.

\section{Filter ghost}\label{ap:fil}
Similarly, the 3rd-order polynomials used to describe the filter ghost offsets (from the filter ghost position to the source star position) are also given by \cref{eq:poly}, and can thus be calculated in the same way as the dichroic ghosts described in \cref{ap:di}. The matrices containing the coefficients for the filter ghost offset in each axis, used for the Q1 models, are

\begin{align}
\mathrm{Offset}_{Y_\mathrm{MOSA}} &= 
\begin{pmatrix*}[l]
+1.10450533 & -3.37612818\times10^{-4} & +3.49727260\times10^{-6}\\
+1.42692340\times10^{-1} & -6.90666206\times10^{-7} & +1.41823342\times10^{-7}\\
+1.56788085\times10^{-5} & -3.31387125\times10^{-8} & -9.04039067\times10^{-10}\\
\end{pmatrix*}~\mathrm{and}\\[5mm]
\mathrm{Offset}_{Z_\mathrm{MOSA}} &= 
\begin{pmatrix*}[l]
-2.62240029\times10^{-2} & +1.42325301\times10^{-1} & +1.93865598\times10^{-6}\\
-3.99084565\times10^{-5} & +1.32211002\times10^{-5} & +1.07744649\times10^{-8}\\
+2.56165095\times10^{-6} & +9.63337612\times10^{-8} & -8.67659173\times10^{-10}\\
\end{pmatrix*}.\\[5mm]
\end{align}

In addition to the offset, the ratio between the major and minor
axis (i.e. the elongation) of the filter ghost, as well as the orientation of the ellipse describing the generic shape of the filter ghost also depends on the position on the FPA. The 2nd-order polynomials used to describe these relations are also in the same form as \cref{eq:poly}, and can thus also be calculated using the same method in Python i.e. similarly to \cref{eq:poly_python}. The matrices containing the coefficients used to describe the elongation ($\mathrm{radius_{ratio}}$) and orientation relations, used for the Q1 models, are

\begin{align}
\mathrm{radius_{ratio}} &= 
\begin{pmatrix*}[l]
+1.06612159 & +1.47073999\times10^{-4} & +1.54457330\times10^{-5}\\
+2.35145940\times10^{-4} & +2.77414315\times10^{-6} & \makebox[2.9cm][c]{0}\\
+5.35342973\times10^{-6} & \makebox[2.9cm][c]{0} & \makebox[2.9cm][c]{0}\\
\end{pmatrix*}\mathrm{, and}\\[5mm]
\mathrm{orientation} &= 
\begin{pmatrix*}[l]
-2.01850578\times10^{-2} & -1.63892790\times10^{-3} & +3.96230948\times10^{-6}\\
-3.39665214\times10^{-5} & -1.70575549\times10^{-4} & \makebox[2.9cm][c]{0}\\
+2.59390644\times10^{-6} & \makebox[2.9cm][c]{0} & \makebox[2.9cm][c]{0}\\
\end{pmatrix*}.
\end{align}

The minor axis of the filter-ghost mask ($R_{\rm mask_{min}}$) is described by \cref{eq:radius}, and is thus calculated the same way as the dichroic-ghost mask radius with the values given in \cref{table:rad}. The major axis of the filter-ghost mask (using Python methods) is then calculated by

\begin{equation}
    R_{\rm mask_{maj}} = R_{\rm mask_{min}}*\mathrm{numpy.polynomial.polynomial.polyval2d}(Y_\mathrm{MOSA},Z_\mathrm{MOSA},\mathrm{radius_{ratio}}),
\end{equation}

while the orientation of $R_{\rm mask_{maj}}$, moving clockwise from the $Y_\mathrm{MOSA}$ axis in radians, is given by
\begin{equation}
    R_{\rm mask_{orientation}} = \mathrm{numpy.polynomial.polynomial.polyval2d}(Y_\mathrm{MOSA},Z_\mathrm{MOSA},\mathrm{orientation}).
\end{equation}
   
\end{appendix}
\end{document}